\documentclass[prb,amsmath,amssymb,amsbsy,twocolumn,showpacs]{revtex4-1}
\usepackage{graphicx,color,dcolumn,bm}
\begin{document}

\title{Direct manifestation of band topology in the winding number of the Wannier-Stark ladder}

\author{Woo-Ram Lee}
%\affiliation{Quantum Universe Center, Korea Institute for Advanced Study, Seoul 130-722, Korea}
\author{Kwon Park}
\email{kpark@kias.re.kr}
\affiliation{Quantum Universe Center and School of Physics, Korea Institute for Advanced Study, Seoul 130-722, Korea}

\date{\today}

\begin{abstract}
Topological quantum phases of matter have been a topic of intense interest in contemporary condensed matter physics. 
Extensive efforts are devoted to investigate various exotic properties of topological matters including topological insulators, topological superconductors, and topological semimetals. 
For topological insulators, the dissipationless transport via gapless helical edge or surface states is supposed to play a defining role, which unfortunately
has proved difficult to realize in experiments due to inevitable backscattering induced in the sample boundary. 
Motivated by the fundamental connection between topological invariants and the Zak phase, here, we show that the non-trivial band topologies of both two and three-dimensional topological insulators, characterized by the Chern numbers and the $\mathbb{Z}_2$ invariants, respectively, are directly manifested in the winding numbers of the Wannier-Stark ladder (WSL) emerging under an electric field.
We use the Floquet Green's function formalism to show that the winding number of the WSL is robust against interband interference as well as non-magnetic impurity scattering.
\end{abstract}

\maketitle

%%%%%%%%%%%%%
%%% Introduction %%%
%%%%%%%%%%%%%

\section{Introduction}
\label{sec:Introduction}

Formulated in terms of algebraic commutation relations between operators, quantum mechanics had not been usually related with geometry or topology before the discovery of the geometric phase~\cite{Pancharatnam56, Berry84}, or more commonly known as the Berry phase.
The notion that the topological structure of the Berry phase can be used as a new ``order parameter'' distinguishing between different quantum phases of matter has triggered an intense outburst of research activities in contemporary condensed matter physics~\cite{Hasan10, Qi11, Hohenadler13, Witczak-Krempa14}.

In two dimensions, where perpendicular spin components are conserved, such an order parameter is the spin-dependent Chern number, which can be in principle measured via the spin Hall conductance according to the Kubo formula~\cite{Qi06a,Qi06b}:
$G_\mathrm{H}^\mathrm{spin} = (\mathcal{C}_\uparrow - \mathcal{C}_\downarrow) e^2/h$,
where $\mathcal{C}_\sigma$ is the Chern number of an occupied band with spin component $\sigma$. 
Unfortunately, fully spin-filtered measurements are very difficult to perform in experiments.
An alternative is to measure the ballistic two-terminal charge conductance, which is observed to be quantized approximately as $G=2e^2/h$ in a two-dimensional (2D) topological insulator (TI)~\cite{Konig07, Konig08, Roth09} agreeing with a theoretical prediction assuming that gapless helical edge states generate dissipationless charge transport~\cite{Kane05b}.

In contrast to chiral edge states in the quantum Hall effect, however, the helical edge states are inevitably coupled to various backscattering sources induced in the sample boundary so that the conductance quantization is not exactly protected~\cite{Wu06, Xu06, Maciejko09,  Schmidt12, Cheianov13, Altshuler13, Vayrynen13, Vayrynen14}. 
This means that the edge transport measurement is not an ideal method to reveal the bulk band topology of 2D TIs, which remains to be protected against non-magnetic impurity scattering as well as other perturbations. 
Given this problem, it is beneficial to devise a physical observable directly manifesting the topological order in the bulk without reference to the non-universal electron dynamics in the sample boundary.
In this context, it is worth mentioning that the spin-charge separation in the presence of a $\pi$ flux defect can be used as a possible avenue to reveal the band topology in the bulk~\cite{Qi08a, Ran08}.

The situation becomes even more complicated in three dimensions, where different spin components are in general mixed and thus the spin-dependent Chern number is not properly defined.
In this situation, proper topological order parameters are the $\mathbb{Z}_2$ invariants, $(\nu_0;\nu_1,\nu_2,\nu_3)$, which can fully characterize the three-dimensional (3D) band topology if both inversion and time-reversal symmetries are present.  
Unfortunately, no simple transport measurement can manifest these 3D topological invariants as directly as the (spin Hall or two-terminal charge) conductance in two dimensions. 
Instead, the strong $\mathbb{Z}_2$ topological invariant $\nu_0$ (most important for the robustness of a given 3D TI) is predicted to be manifested in the quantized magneto-electric effect, where an electric field induces a topological contribution to the magnetization~\cite{Qi08b}.   
While an actual measurement of the magneto-electric effect may be too difficult to perform at present, the fact that the topological invariant has a direct physical consequence is important as a matter of principle.

In practice, the 3D band topology has been inferred from the existence of helical surface states, which exhibit the spin-momentum locking as a consequence of the non-trivial topological order~\cite{Hsieh08, Xia09}. 
However, a problem is that the spin-momentum locking by itself is not directly related with any of the 3D topological invariants.    
In some sense, it is even more beneficial in three dimensions to devise a physical observable directly manifesting the topological invariants (in addition to the magneto-electric effect).
%Also, the helical surface states are subject to various backscattering sources induced in the sample boundary similar to the helical edge states in two dimensions. 

Motivated by the fundamental connection between topological invariants and the Zak phase, in this work, we show that the non-trivial band topologies of both 2D and 3D TIs (characterized by the Chern numbers and the $\mathbb{Z}_2$ invariants, respectively) are directly manifested in the energy spectrum of electrons under an electric field via the winding number of the Wannier-Stark ladder (WSL). 
The WSL is a set of energy eigenstates of electrons confined in the lattice under an electric field, which are the quantized modes of the Bloch oscillation.
In contrast to a recent interferometric method proposed in optical lattices~\cite{Abanin13, Liu13, Dauphin13, Grusdt14}, which combines the coherent Bloch oscillation with the Ramsey interferometry, our spectroscopic method can be applied to condensed matter systems, where the phase coherence is not guaranteed.  
Concretely, we use the Floquet Green's function formalism to show that the winding number of the WSL is robust against interband interference as well as non-magnetic impurity scattering.

Provided that the fully interacting Floquet Green's function is obtained, our method can be applied to any strongly correlated systems with the non-trivial band topology in order to address various theoretical issues such as (i) how far the topological order can persist as a function of correlation strength and (ii) what new phases of strongly correlated topological matter can emerge at sufficiently strong correlation. 
It is worth mentioning that the definitions for topological invariants were previously extended to general strongly correlated systems in equilibrium, i.e., without the electric field~\cite{Wang10_PRL, Wang12, Go12}. 
Experimentally, this means that the magneto-electric effect (which is directly related with a topological invariant) can be used to characterize strongly correlated topological insulators. 
In our theory, the non-trivial winding number of the WSL can play a similar role as the magneto-electric effect.

The rest of the paper is organized as follows.
In Sec.~\ref{sec:Band_topology}, we summarize briefly how the 2D and 3D band topologies are characterized by the Chern numbers and the $\mathbb{Z}_2$ invariants, respectively.
In Sec.~\ref{sec:Winding_number}, we explain how the winding number of the WSL can provide a direct manifestation of topological invariants. 
Specifically, our theory is presented in three levels of complication; 
(i) the semiclassical theory of the Abelian Berry connection/curvature, which can be applied to 2D TIs in the adiabatic limit (Sec.~\ref{sec:Semiclassical_theory}),
(ii) the requantized effective theory of the general non-Abelian Berry connection/curvature, which can be applied to 2D as well as 3D TIs in the adiabatic limit (Sec.~\ref{sec:Requantized_effective_theory}), and 
(iii) the full quantum theory using the Floquet Green's function formalism, which can be applied to general situations (Sec.~\ref{sec:Full_quantum_theory}).
We present computational results in Sec.~\ref{sec:Results}, proving that the winding number of the WSL provides a direct manifestation of topological invariants in both 2D and 3D TIs.
We conclude in Sec.~\ref{sec:Discussion}, where we discuss the experimental feasibility of observing the winding number of the WSL.

%%%%%%%%%%%%%%
%%% Band topology %%%
%%%%%%%%%%%%%%
\section{Band topology}
\label{sec:Band_topology}

%%%%%%%%%%%%%%%
%% Chern numbers in 2D %%
%%%%%%%%%%%%%%%
\subsection{Chern numbers in two dimensions}
\label{sec:Chern_numbers}

Let us begin with 2D TIs, where perpendicular spin components are conserved. 
The Hamiltonian for 2D TIs has the following generic structure~\cite{Hasan10, Qi11, Bernevig06, Kane05a}: 
$H = \sum_{\mathbf{k},\sigma=\uparrow,\downarrow} \psi_{\mathbf{k}\sigma}^\dag H_\sigma(\mathbf{k}) \psi_{\mathbf{k}\sigma}$ with
$H_\downarrow(\mathbf{k}) = H_\uparrow^*(-\mathbf{k})$ and
\begin{align}
H_\uparrow(\mathbf{k})
= \epsilon_\mathbf{k} \mathbb{I}_2 + \mathbf{d}_\mathbf{k} \cdot \boldsymbol{\sigma} 
= \left(
\begin{array}{cc}
\epsilon_\mathbf{k} + d_{\mathbf{k},z}
& d_{\mathbf{k},-} \\
d_{\mathbf{k},+}
& \epsilon_\mathbf{k} - d_{\mathbf{k},z}
\end{array}
\right),
\label{Hamiltonian2DTI}
\end{align}
where $\psi_{\mathbf{k}\sigma}^\dag = (c_{\mathbf{k} \alpha \sigma}^\dag, c_{\mathbf{k} \alpha^\prime \sigma}^\dag)$ 
with $c_{\mathbf{k}\alpha\sigma}^\dag$ being the electron creation operator with momentum $\mathbf{k}$, spin $\sigma$, and orbital $\alpha$. 
The physical meaning of $\alpha$ depends on the specific model.
Concretely, %with $\alpha^\prime$ indicating the opposite orbital of $\alpha$, 
$\alpha$ denotes the sublattice indices, $A$ or $B$, in the Kane-Mele (KM) model, and the conduction/valence band indices, $E_1$ or $H_1$, in the Bernevig-Hughes-Zhang (BHZ) model. 
$\mathbb{I}_n$ is the $n\times n$ identity matrix.
$\boldsymbol{\sigma} = (\sigma_x, \sigma_y, \sigma_z)$ consists of the Pauli matrices.
The $\mathbf{d}_\mathbf{k}$ vector has three components $\mathbf{d}_\mathbf{k} = (d_{\mathbf{k},x}, d_{\mathbf{k},y}, d_{\mathbf{k},z})$, where the first two components can be combined as $d_{\mathbf{k},\pm} = d_{\mathbf{k},x} \pm i d_{\mathbf{k},y}$.
The band dispersion is given by $\mathcal{E}_\pm (\mathbf{k}) = \epsilon_\mathbf{k} \pm |\mathbf{d}_\mathbf{k}|$ and $\epsilon_\mathbf{k} \pm |\mathbf{d}_{-\mathbf{k}}|$ for spin up and down, respectively, indicating that the system becomes insulating when $|\mathbf{d}_\mathbf{k}| \neq 0$.

The 2D band topology is characterized by the total flux of the Berry curvature piercing through the entire Brillouin zone (BZ) for each spin component.
Called the (first) Chern number, the total flux of the Berry curvature is
equivalent to the wrapping number of the normalized $\mathbf{d}_\mathbf{k}$ field around the unit sphere,
\begin{align}
\mathcal{C}_\uparrow = - \mathcal{C}_\downarrow =  \frac{1}{4\pi} \int_{\rm BZ} dk_x dk_y~\hat{\mathbf{d}}_\mathbf{k} \cdot ( \partial_{k_x} \hat{\mathbf{d}}_\mathbf{k} \times \partial_{k_y} \hat{\mathbf{d}}_\mathbf{k} ),
\label{Chern_number_definition}
\end{align}
where $\hat{\mathbf{d}}_\mathbf{k} = \mathbf{d}_\mathbf{k} / |\mathbf{d}_\mathbf{k}|$~\cite{Hasan10,Qi11}. 
In fact, the 2D band topology can be fully determined by examining the low-energy behavior of $\mathbf{d}_{\mathbf{k}=\mathbf{K}+\mathbf{q}}$ around $\mathbf{K} = 0$ or other low-energy momenta, where $\mathbf{d}_{\mathbf{K}+\mathbf{q}}$ can be generally expanded as $(A q_x, \pm A q_y, M + B (q_x^2 + q_y^2))$.
When a single low-energy point exists at $\mathbf{K}=0$~\cite{Bernevig06}, the band topology becomes non-trivial if $M/B < 0$ and trivial otherwise. 
In the Kane-Mele model~\cite{Kane05a, Kane05b} defined on the honeycomb lattice, there are two low-energy Dirac points at $\mathbf{K}=\mathbf{K}^{\pm}=(\frac{2\pi}{3a}, \pm \frac{2\pi}{3\sqrt{3}a})$, both of which should satisfy the non-triviality condition in order for the whole valence band to become topologically non-trivial.

In the presence of the time-reversal symmetry, the spin-dependent Chern numbers are always opposite between different spin components, which means that the 2D band topology is fully characterized by the Chern number difference.
Motivated by the analogy between the charge and time-reversal polarization (TRP), the Chern number difference can be alternatively computed in a discrete form, which is formulated in terms of the parities of the time-reversal operator, $\delta_{i(=1,2,3,4)}$, at four time-reversal invariant momenta (TRIM)~\cite{Kane05a}:
\begin{align}
(-1)^{\nu_{\rm 2D}} = \prod_{i=1}^4 \delta_i ,
\label{2D_Z2_invariant}
\end{align}
where the 2D $\mathbb{Z}_2$ invariant, $\nu_{\rm 2D}$, is identical to the half of the Chern number difference computed in an integral form in Eq.~\eqref{Chern_number_definition}: 
\begin{align}
\nu_{\rm 2D} = \frac{\mathcal{C}_\uparrow - \mathcal{C}_\downarrow}{2}~~\textrm{(mod 2)}.
\end{align}
As shown in the following section, the fact that the 2D topological invariant can be computed in a discrete form has played an important role in defining the 3D topological invariants.

%%%%%%%%%%%%%%
%% Z2 invariants in 3D %%
%%%%%%%%%%%%%%
\subsection{$\mathbb{Z}_2$ invariants in three dimensions}
\label{sec:Z2_invariants}

In three dimensions, different spin components are in general mixed.
The Hamiltonian for 3D TIs has the following generic structure~\cite{Hasan10, Qi11}: 
$H = \sum_{\mathbf{k}} \psi_\mathbf{k}^\dag H(\mathbf{k}) \psi_\mathbf{k}$ with
\begin{align}
H(\mathbf{k})
& = \epsilon_\mathbf{k} \mathbb{I}_4 + \mathbf{d}_\mathbf{k} \cdot \boldsymbol{\Gamma} \nonumber \\
& = \left(
\begin{array}{cccc}
\epsilon_\mathbf{k} - d_{\mathbf{k},3} & d_{\mathbf{k},4} & 0 & d_{\mathbf{k},-} \\
d_{\mathbf{k},4} & \epsilon_\mathbf{k} + d_{\mathbf{k},3} & d_{\mathbf{k},-} & 0 \\
0 & d_{\mathbf{k},+} & \epsilon_\mathbf{k} - d_{\mathbf{k},3} & - d_{\mathbf{k},4} \\
d_{\mathbf{k},+} & 0 & - d_{\mathbf{k},4} & \epsilon_\mathbf{k} + d_{\mathbf{k},3}
\end{array}
\right),
\label{Hamiltonian3DTI}
\end{align}
where $\psi_{\bf k}^\dag = (c_{{\bf k}\alpha_1}^\dag, c_{{\bf k}\alpha_2}^\dag, c_{{\bf k}\alpha_3}^\dag, c_{{\bf k}\alpha_4}^\dag)$ with $c_{{\bf k}\alpha}^\dag$ being the electron creation operator with momentum ${\bf k}$ on generalized orbital $\alpha$, which includes both spin and orbital degrees of freedom.
As before, the physical meaning of $\alpha$ depends on the specific model.
In the model for BiSe-family materials, $(\alpha_1, \alpha_2, \alpha_3, \alpha_4) = ( P1_z^+\!\uparrow, P2_z^-\!\uparrow, P1_z^+\!\downarrow, P2_z^-\!\downarrow )$. 
$\mathbb{I}_n$ is the $n\times n$ identity matrix. 
$\boldsymbol{\Gamma} = (\Gamma_1, \Gamma_2, \Gamma_3, \Gamma_4, \Gamma_5)$ consists of the Gamma matrices satisfying the Clifford algebra $\{\Gamma_i,\Gamma_j\} = 2\delta_{ij}\mathbb{I}_4$. 
%Also, $\mathbf{d}_\mathbf{k} = (d_{\mathbf{k},1}, d_{\mathbf{k},2}, d_{\mathbf{k},3}, d_{\mathbf{k},4}, d_{\mathbf{k},5})$ with $d_{{\bf k},\pm}=d_{{\bf k},1} \pm i d_{{\bf k},2}$. 
Since there are five Gamma matrices, the corresponding ${\bf d}_{\bf k}$ vector has also five components; 
$\mathbf{d}_\mathbf{k} = (d_{\mathbf{k},1}, d_{\mathbf{k},2}, d_{\mathbf{k},3}, d_{\mathbf{k},4}, d_{\mathbf{k},5})$. 
Specifically, for BiSe-family materials, ${\bf d}_{\bf k}$ can be expanded around the $\Gamma$ point as 
$d_{{\bf k},\pm}=d_{{\bf k},1} \pm i d_{{\bf k},2} \simeq A_1 (k_x \pm i k_y)$,
$d_{{\bf k},3} \simeq M+B_1(k_x^2+k_y^2)+B_2 k_z^2$,
$d_{{\bf k},4} \simeq A_2 k_z$, and
$d_{{\bf k},5} \simeq 0$.
Similarly, $\epsilon_{\bf k}$ can be expanded as $\epsilon_\mathbf{k} \simeq C + D_1 (k_x^2+k_y^2) +D_2 k_z^2$.

In the presence of mixing between different spin components, the spin-dependent Chern numbers cannot be defined properly in 3D TIs.
In fact, mathematically, the Chern number cannot be defined at all in three dimensions. 
Fortunately, if both inversion and the time-reversal symmetries are present, the 3D band topology can be characterized by four $\mathbb{Z}_2$ invariants, $(\nu_0;\nu_1,\nu_2,\nu_3)$, which depend on the parities of the time-reversal operator, $\delta_{i (=1,\cdots,8)},$  at eight TRIM~\cite{Fu07, Fu07b}. 
Playing the most important role by governing the robustness of a given 3D TI, the strong $\mathbb{Z}_2$ invariant, $\nu_0$, is defined as 
\begin{align}
(-1)^{\nu_0} 
= \prod_{i=1}^8 \delta_i 
%= \prod_{i=1}^4 \delta_i \prod_{j=5}^8 \delta_j 
= (-1)^{\nu_{\rm 2D}} (-1)^{\nu_{{\rm 2D}^\prime}},
\label{3D_Z2_invariant}
\end{align}
where $\nu_{\rm 2D}$ and $\nu_{{\rm 2D}^\prime}$ are the 2D $\mathbb{Z}_2$ invariants of the inversion-symmetric 2D subspaces containing one set of four TRIM and the other, respectively~\cite{Grusdt14}.
It is worth mentioning that $\nu_0$ is proportional to the integral of the Chern-Simons three form over the 3D BZ, which is only quantized in the presence of both inversion and the time-reversal symmetries~\cite{Wang10}.

Equation~\eqref{3D_Z2_invariant} implies that the 3D TI becomes a strong TI if and only if the band topology of the inversion-symmetric 2D subspace containing one set of four TRIM is opposite to that of the 2D subspace containing the other.
In the cubic lattice, this means that the band topology of the $k_z a=0$ subspace should be topologically non-trivial if that of the $k_z a=\pi$ subspace is topologically trivial, and vice versa.  
Of course, this statement should be true regardless of whether we choose the $k_x a= 0$ and  $\pi$ subspaces (or the $k_y a=0$ and $\pi$ subspaces) instead of the $k_z a=0$ and $\pi$ counterparts.
In this example, the other three $\mathbb{Z}_2$ invariants simply correspond to the 2D $\mathbb{Z}_2$ invariants of the $k_x=0$, $k_y=0$, and $k_z=0$ subspaces:
that is to say, $\nu_1=\nu_{{\rm 2D}(k_x=0)}$, $\nu_2=\nu_{{\rm 2D}(k_y=0)}$, and $\nu_3=\nu_{{\rm 2D}(k_z=0)}$.

%%%%%%%%%%%%%%%%%%%%%
%%% Winding number of the WSL %%%
%%%%%%%%%%%%%%%%%%%%%
\section{Winding number of the Wannier-Stark ladder}
\label{sec:Winding_number}

%%%%%%%%%%%%%%%
%% Semiclassical theory %%
%%%%%%%%%%%%%%%
\subsection{Semiclassical theory of the Abelian Berry connection/curvature}
\label{sec:Semiclassical_theory}

%%%%%%%%%%%%%%%%%%%%%%%%%%%%%%%%%%%%%%%%%%%%%%%%%%%%%%%%%%%%%%%%%%%%%
%%%%%%%%%%%%%%%%%%%%%%%%%%%%%%%%%%%%%%%%%%%%%%%%%%%%%%%%%%%%%%%%%%%%%
\begin{figure*}[t]
\centering
\includegraphics[width=0.9\textwidth]
{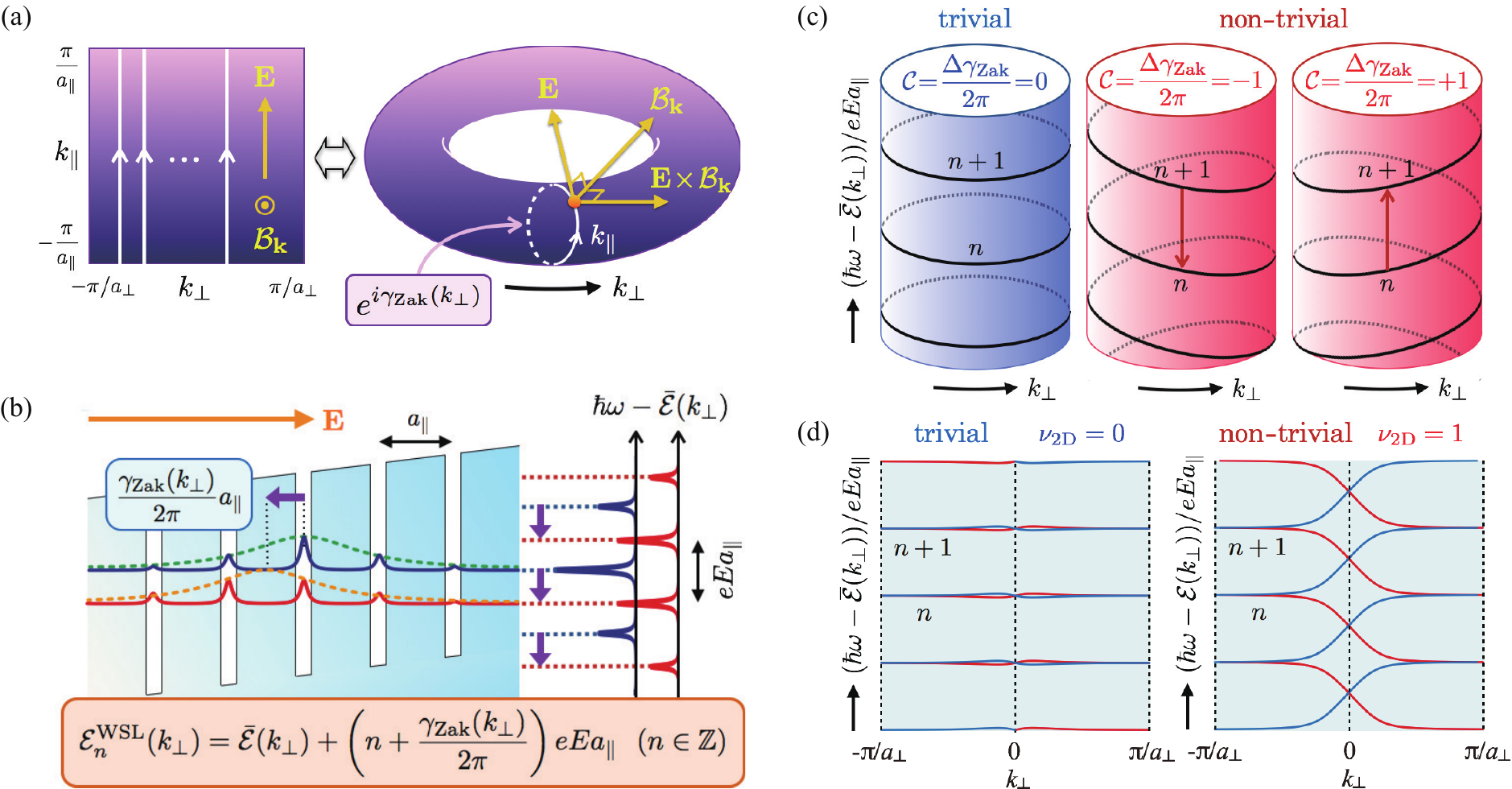} \\
\caption{
Schematic diagrams illustrating the fundamental connection between the 2D topological invariant and the winding number of the WSL.
Figure~\ref{Fig1_Abelian_schematic}~(a) depicts the semiclassical situation, where an electron wave packet performs the Bloch oscillation in the 2D momentum space under an electric field $\mathbf{E}$. 
In the presence of the Abelian Berry curvature $\mathcal{B}_\mathbf{k}$, the Bloch oscillation is affected by the anomalous velocity $e \mathbf{E} \times \mathcal{B}_\mathbf{k} /\hbar$. 
The semiclassical path of the Bloch oscillation (encircling the torus along the $k_\parallel$ direction) is quantized via the Bohr-Sommerfeld quantization rule with an additional geometrical factor $e^{i\gamma_{\rm Zak}(k_\perp)}$, where $\gamma_{\rm Zak}(k_\perp)$ is the Zak phase.
Figure~\ref{Fig1_Abelian_schematic}~(b) shows that the center positions of the WSL eigenstates (i.e., the center positions of the envelope wave functions plotted via green and orange dashed lines) are shifted by $\gamma_\mathrm{Zak}(k_\perp) a_\parallel /2\pi$, which generates  a corresponding shift in the energy spectrum of the WSL eigenstates. 
For clarity, only the wave function for the central WSL branch is shown. 
Figure~\ref{Fig1_Abelian_schematic}~(c) illustrates that the trivial/non-trivial Chern number is directly manifested in the trivial/non-trivial winding number of the WSL as a function of $k_\perp$. 
Figure~\ref{Fig1_Abelian_schematic}~(d) shows that, with the time-reversal symmetry dictating that the spin-dependent Chern numbers are opposite between different spin components, i.e., $\mathcal{C}_\uparrow=-\mathcal{C}_\downarrow$, two separate sets of the spin-dependent WSL branches (distinguished by red and blue lines) wind oppositely in the topologically non-trivial phase, which is characterized by the 2D $\mathbb{Z}_2$ invariant $\nu_{\rm 2D}=(\mathcal{C}_\uparrow-\mathcal{C}_\downarrow)/2=1$ (mod 2).
In the topologically trivial phase characterized by $\nu_{\rm 2D}=0$, there is no winding of the WSL.
Note that the guide lines for the WSL branches are obtained from the actual solution of the semiclassical theory for the Bernevig-Hughes-Zhang model via Eq.~\eqref{WSL_spectrum}.
}
\label{Fig1_Abelian_schematic}
\end{figure*}
%%%%%%%%%%%%%%%%%%%%%%%%%%%%%%%%%%%%%%%%%%%%%%%%%%%%%%%%%%%%%%%%%%%%%
%%%%%%%%%%%%%%%%%%%%%%%%%%%%%%%%%%%%%%%%%%%%%%%%%%%%%%%%%%%%%%%%%%%%%

To appreciate how the energy spectrum of electrons under an electric field can reveal the band topology, it is instructive to first consider the semiclassical dynamics of an electron wave packet moving in the lattice. 
Here, for simplicity, we assume that the electron wave packet is entirely composed of the plane waves consisting in a single nondegenerate band, which is well separated from other bands in the full energy spectrum. 
Also, for the time being, let us focus on the 2D TI, where perpendicular spin components are conserved so that the Hamiltonian for each spin component is decoupled.
In this situation, the Berry connection/curvature becomes Abelian. 
An extension to the general case of the non-Abelian Berry connection/curvature is discussed in Sec.~\ref{sec:Requantized_effective_theory}.

Under an electric field, the electron wave packet in the lattice performs the Bloch oscillation~\cite{Bloch28}, whose dynamics is described by the semiclassical Lagrangian~\cite{Sundaram99} 
\begin{align}
\mathcal{L}(\mathbf{r},\dot{\mathbf{r}},\mathbf{k},\dot{\mathbf{k}}) = \hbar\mathbf{k}\cdot\dot{\mathbf{r}} + \hbar\mathcal{A}_n(\mathbf{k})\cdot\dot{\mathbf{k}} -\mathcal{H}_n(\mathbf{r},\mathbf{k}),
\label{semi_Lag1}
\end{align}
where ${\bf r}$ is the center position, $\hbar{\bf k}$ is the mean crystal momentum, and 
$\mathcal{H}_n(\mathbf{r},\mathbf{k}) = \mathcal{E}_n(\mathbf{k}) + e\mathbf{E} \cdot \mathbf{r}$ is the semiclassical Hamiltonian for the $n$-th energy band. 
$\mathcal{A}_n (\mathbf{k}) = \langle \phi_n(\mathbf{k}) |i\nabla_\mathbf{k} |\phi_n(\mathbf{k}) \rangle$ is the Berry connection with $\phi_n(\mathbf{k})$ being the periodic part of the Bloch wave function for the $n$-th energy band. 
In the case of the two-band model in Eq.~\eqref{Hamiltonian2DTI}, there are two degenerate energy bands for each spin component, $\mathcal{E}_{\pm,\sigma}(\mathbf{k})$, and the corresponding Berry connections, $\mathcal{A}_{\pm,\sigma}(\mathbf{k})$.
From this forward, we focus on the lower occupied band with spin up so that we simplify the notation by setting $\mathcal{H}(\mathbf{r},\mathbf{k}) = \mathcal{H}_{-,\uparrow}(\mathbf{r},\mathbf{k})$, $\mathcal{E}_\mathbf{k}=\mathcal{E}_{-,\uparrow}(\mathbf{k})$, and $\mathcal{A}_\mathbf{k}=\mathcal{A}_{-,\uparrow}(\mathbf{k})$. 
Note that the same analyses presented below can be repeated for spin down.

By applying the variational method to the above Lagrangian, one can derive two equations of motion. %for $\mathbf{k}$ and $\mathbf{r}$.
First, $\hbar \dot{\mathbf{k}} = -e\mathbf{E}$, which tells us that the crystal momentum parallel to the electric field, $\hbar k_\parallel$, changes linearly in time, while the perpendicular crystal momentum, $\hbar k_\perp$, remains fixed.
Here, the charge of electron is defined to be $-e$.
Second, $\hbar \dot{\mathbf{r}} = \nabla_\mathbf{k} \mathcal{E}_\mathbf{k}+e\mathbf{E} \times \mathbf{\mathcal{B}}_\mathbf{k}$, where
the former term of the right-hand side is the usual group velocity and the latter is the anomalous velocity due to the (Abelian) Berry curvature $\mathbf{\mathcal{B}}_\mathbf{k} = \nabla_\mathbf{k} \times \mathcal{A}_\mathbf{k}$.
Without the anomalous velocity, these equations of motion describe the usual Bloch oscillation, which can be understood in terms of the Bragg scattering of a wave packet at the BZ boundaries.

The anomalous velocity generates not only a bending of the Bloch-oscillation orbit, but also a shift of its center. 
Considering that the quantized mode of the Bloch-oscillation orbit is nothing but the WSL eigenstate~\cite{Mendez93, Gluck02}, this means that the WSL eigenstate centers are shifted by the Berry curvature effect. 
To see how this is possible, let us subtract a total derivative $\hbar \frac{d}{dt} (\mathbf{k}\cdot\mathbf{r})$ from the Lagrangian in Eq.~\eqref{semi_Lag1}, which generates the following new Lagrangian
\begin{align}
\mathcal{L}^\prime(\mathbf{r},\mathbf{k},\dot{\mathbf{k}}) = -\hbar\mathcal{R}(\mathbf{r}, \mathbf{k}) \cdot \dot{\mathbf{k}} -\mathcal{H}(\mathbf{r},\mathbf{k}),
\label{semi_Lag2}
\end{align}
where $\mathcal{R}(\mathbf{r},\mathbf{k})= \mathbf{r} - \mathcal{A}_\mathbf{k}$ can be interpreted as the canonical {\it position} conjugate to $\mathbf{k}$.
In this interpretation, the Bloch-oscillation orbit is quantized according to the Bohr-Sommerfeld quantization rule:
\begin{align}
\oint_C d\mathbf{k}_\parallel \cdot \mathcal{R}(\mathbf{r}, \mathbf{k}) = 2\pi n  \;\;\; (n \in \mathbb{Z}) ,
\end{align}
where $C$ denotes a closed orbit with the constant energy, where $k_\parallel a_\parallel$ sweeps through the entire BZ between $-\pi$ and $\pi$ with $k_\perp a_\perp$ fixed.
Here, $a_\parallel$ is the lattice constant of the projected unit cell along the parallel direction to the electric field. 
Meanwhile, $a_\perp$ is defined so that $a_\parallel a_\perp$ is the area of the unit cell.

As a consequence of the Bohr-Sommerfeld quantization rule, the center position, or the {\it polarization} of the WSL eigenstates is quantized according to 
$\oint_C d{\bf k}_\parallel \cdot {\bf r} = 2\pi n + \gamma_\mathrm{Zak}(k_\perp)$, 
where the non-integer shift is called the Zak phase~\cite{Zak89, King-Smith93, Delplace11}:
\begin{align}
\gamma_\mathrm{Zak}(k_\perp)
%= \oint_C dk_\parallel \mathcal{A}_{\mathbf{k}, \parallel} , 
= \oint_C d{\bf k}_\parallel \cdot \mathcal{A}_\mathbf{k} , 
\label{Zak_phase}
\end{align}
which is generally a function of $k_\perp$. 
%%%%%%%%%%%%%%%%%%%%% Reference to Appendix A %%%%%%%%%%%%%%%%%%%%%%%%%%%%%%%%%%%%
See Appendix~\ref{Appen:Zak} for computational details of $\gamma_\mathrm{Zak}(k_\perp)$.
%%%%%%%%%%%%%%%%%%%%%%%%%%%%%%%%%%%%%%%%%%%%%%%%%%%%%%%%%%%%%%%%%%%%%%
Now, by equating the energy of the WSL eigenstates with the averaged semiclassical Hamiltonian over $C$, $\frac{a_\parallel}{2\pi} \oint_C dk_\parallel \mathcal{H}(\mathbf{r}, \mathbf{k})$, 
one can show that the energy spectrum of the WSL eigenstates is given by
\begin{align}
\mathcal{E}^{\rm WSL}_n(k_\perp)
& = \bar{\mathcal{E}}(k_\perp) + \left( n + \frac{\gamma_\mathrm{Zak}(k_\perp)}{2\pi} \right) eE a_\parallel , 
\label{WSL_spectrum}
\end{align}
where $\bar{\mathcal{E}}(k_\perp)=\frac{a_\parallel}{2\pi} \oint_C dk_\parallel \mathcal{E}_\mathbf{k}$.  %is the mean energy with respect to $k_\parallel$. 

The Zak phase is related with the Chern number, i.e., the total flux of the (Abelian) Berry curvature piercing through the entire BZ,
\begin{align}
\mathcal{C} = \frac{1}{2\pi} \int_\mathrm{BZ} d^2\mathbf{k} \cdot \mathbf{\mathcal{B}}_\mathbf{k}, 
\end{align}
which, by using the Stoke's theorem, can be rearranged as follows~\cite{King-Smith93}: 
\begin{align}
\mathcal{C}
=\int^{\pi/a_\perp}_{-\pi/a_\perp} \frac{dk_\perp}{2\pi} \frac{\partial \gamma_\mathrm{Zak}(k_\perp)}{\partial k_\perp} 
= \frac{\Delta \gamma_\mathrm{Zak}}{2\pi} .
\label{RelationshipChernZak}
\end{align} 
Equation~\eqref{RelationshipChernZak} indicates that, if $\mathcal{C}=\pm1$, the WSL index $n$ goes to $n\pm1$ after $k_\perp a_\perp$ sweeps through the entire BZ.  
This means that the Chern number is equivalent to the winding number of the WSL across the BZ as a function of $k_\perp$. 
Figure~\ref{Fig1_Abelian_schematic}~(a)\mbox{--}(c) provide schematic diagrams summarizing the discussions so far.

With both spin components taken into account, the time-reversal symmetry dictates that ${\cal C}_\uparrow= -{\cal C}_\downarrow$.
This means that two separate sets of the WSL branches for spin up and down should wind oppositely and cross each other at $k_\perp a_\perp=0$ and $\pm\pi$, which are TRIM. 
Moreover, following the similar logic predicting that the Kramers doublets should exchange partners in the helical edge states~\cite{Kane05a}, one can predict that the WSL branches should also exchange their Kramers-doublet partners (i.e., the doubly degenerate WSL eigenstates at TRIM) in the topologically non-trivial phase, while not in the trivial phase. 
This prediction is fully confirmed, as shown in Fig.~\ref{Fig1_Abelian_schematic}~(d) and Sec.~\ref{sec:Results}.

%In the next section, we discuss how to deal with the general non-Abelian Berry connection/curvature. 
Finally, it is worthwhile to mention that the Zak phase was previously used as a topological invariant to predict the existence of edge states based on the bulk-edge correspondence between the quantized value of the Zak phase and the existence of a localized edge state~\cite{Delplace11}.
In 2D TIs, the Zak phase is not generally quantized except at TRIM (or equivalently inversion-symmetric momenta), where it becomes either $0$ or $\pi$.   
One of the main points in our work is that the proper topological invariant characterizing the entire 2D band topology is not the Zak phase itself, but rather the change of the Zak phase across the one-dimensional BZ of $k_\perp$, i.e., the winding number of the WSL.

%%%%%%%%%%%%%%%%%%%
%% Requantized effective theory %%
%%%%%%%%%%%%%%%%%%%
\subsection{Requantized effective theory of the general non-Abelian Berry connection/curvature}
\label{sec:Requantized_effective_theory}

We now consider the general case, where the Berry connection/curvature is non-Abelian~\cite{Wilczek84}.
To this end, it is convenient to concentrate on the generic four-band model discussed in Eq.~\eqref{Hamiltonian3DTI} for 3D TIs, which has four energy bands in total with the lower two being degenerate and separated from the upper two (also degenerate) bands by an energy gap.
As before, for simplicity, we focus on the lower two degenerate energy bands, both of which are fully occupied at half filling. 
Generally, in the absence of conserved (pseudo)spin components, the Berry connection/curvature has off-diagonal terms between different energy bands.
Mathematically, the Berry connection in general has the SU(2) non-Abelian gauge structure with $\mathcal{A}_{{\bf k},\alpha\beta} = \langle \phi_{-,\alpha}({\bf k})| i \nabla_{{\bf k}} |\phi_{-,\beta}({\bf k})\rangle$, where $\phi_{-,\alpha}(\mathbf{k})$ and $\phi_{-,\beta}(\mathbf{k})$ are the periodic parts of the Bloch wave function in the lower two degenerate energy bands with $\alpha$ and $\beta$ denoting the pseudospin indices, say, $u$ and $d$.
In the non-Abelian case, the Berry curvature is defined as $\mathcal{B}_{\bf k} = \nabla_{\bf k} \times \mathcal{A}_{\bf k} - i \mathcal{A}_{\bf k} \times \mathcal{A}_{\bf k}$, where the second term does not in general vanish due to the non-commutative relationship between the Berry connections with different spatial coordinates.

The semiclassical Lagrangian can be extended in the non-Abelian case as follows~\cite{Shindou05, Culcer05, Chang08, Xiao10}: 
\begin{align}
\mathcal{L}({\bf r}, \dot{{\bf r}}, {\bf k}, \dot{{\bf k}}, \boldsymbol{\eta}, \dot{\boldsymbol{\eta}}) =& i\hbar \boldsymbol{\eta}^\dag \dot{\boldsymbol{\eta}} + \hbar {\bf k} \cdot \dot{{\bf r}} + \hbar (\boldsymbol{\eta}^\dag \mathcal{A}_{{\bf k}} \boldsymbol{\eta}) \cdot \dot{{\bf k}} 
\nonumber \\
&- \mathcal{H}({\bf r}, {\bf k}),
\end{align} 
where $\mathbf{r}$, $\mathbf{k}$, and $\mathcal{H}(\mathbf{r},\mathbf{k}) = \mathcal{E}(\mathbf{k}) + e\mathbf{E} \cdot \mathbf{r}$ are all defined the same as before in Eq.~\eqref{semi_Lag1}. 
A new addition is the $\boldsymbol{\eta}=(\eta_u,\eta_d)^{\rm T}$ variables, which denote the band decomposition of the wave packet satisfying the normalization condition, $\sum_{\alpha=u,d} |\eta_\alpha|^2 = 1$. 
The equation of motion for ${\bf r}$ is obtained similar to the Abelian case with a modification that the Berry curvature $\mathcal{B}_{{\bf k}}$ is now replaced by the $\boldsymbol{\eta}$-averaged non-Abelian Berry curvature $\boldsymbol{\eta}^\dag \mathcal{B}_{{\bf k}} \boldsymbol{\eta}$. 
The dynamics of $\boldsymbol{\eta}$ is governed by $i\hbar \dot{\boldsymbol{\eta}} = e{\bf E} \cdot \mathcal{A}_{\bf k} \boldsymbol{\eta}$.
The equation of motion for ${\bf k}$ is the same as before in the Abelian case.

The Bohr-Sommerfeld quantization rule is no longer applicable in the non-Abelian case. 
Consequently, it is not obvious how to quantize the Bloch oscillation and obtain the energy spectrum of the WSL eigenstates.
To overcome this obstacle, we requantize the semiclassical theory as follows~\cite{Chang08, Xiao10}.
First, we note that the canonical position $\mathcal{R}({\bf r}, {\bf k})$ in Eq.~\eqref{semi_Lag2} can be generalized to be $\mathcal{R}({\bf r}, {\bf k}, \boldsymbol{\eta}) = {\bf r} - \boldsymbol{\eta}^\dag \mathcal{A}_{\bf k} \boldsymbol{\eta}$. 
Then, we promote $\mathcal{R}$ to a quantum variable, $\hat{\cal R}$, which is conjugate to ${\bf k}$. 
Next, we formally eliminate the $\boldsymbol{\eta}$ variables by promoting the Hilbert space of the requantized effective Hamiltonian to be defined in the $\boldsymbol{\eta}$ pseudospinor space, $|\boldsymbol{\eta}\rangle$. 
Finally, by replacing ${\bf r}$ with ${\bf r}\mathbb{I}_2=\hat{\cal R}\mathbb{I}_2+\mathcal{A}_{\bf k}$ in $\mathcal{H}({\bf r}, {\bf k})$, we obtain the requantized effective Hamiltonian
\begin{align}
\mathcal{H}_{\rm eff}({\bf k}) 
= \mathcal{E}_{{\bf k}} \mathbb{I}_2 + e{\bf E} \cdot (\hat{\cal R} \mathbb{I}_2 + \mathcal{A}_{{\bf k}}),
\label{Heff}
\end{align}
where $\hat{\cal R}= i\nabla_{\bf k}$ since it is conjugate to ${\bf k}$.

The energy spectrum of the WSL eigenstates can be obtained by solving the eigenvalue equation of $\mathcal{H}_{\rm eff}({\bf k})$:
\begin{align}
\mathcal{H}_{\rm eff}({\bf k}) | \boldsymbol{\eta} \rangle
= \mathcal{E}^{\rm WSL}({\bf k}_\perp) | \boldsymbol{\eta} \rangle,
\label{Heff_eigenvalue_eq}
\end{align}
where the WSL eigenenergy, $\mathcal{E}^{\rm WSL}({\bf k}_\perp)$, is shown to be a function of the perpendicular momentum to the electric field, ${\bf k}_\perp$. 
Note that the parallel momentum $k_\parallel$ is not a good quantum number.  
Now, it is convenient to utilize the periodicity in the momentum space and perform the Fourier transformation with respect to $k_\parallel$, which leads to the Floquet-type representation of the eigenvalue equation: 
\begin{align}
\sum_{\beta m} \left[ \mathcal{M}_{\alpha\beta}^{nm}({\bf k}_\perp) +neEa_\parallel \delta_{\alpha\beta} \delta_{nm} \right] \eta_{\beta m} 
= \mathcal{E}^{\rm WSL}({\bf k}_\perp) \eta_{\alpha n},
\label{Heff_Floquet_eq}
\end{align}
where 
\begin{align}
\mathcal{M}_{\alpha\beta}^{nm}({\bf k}_\perp) &= \frac{a_\parallel}{2\pi} \int_0^{\frac{2\pi}{a_\parallel}} dk_\parallel e^{i (n-m) k_\parallel a_\parallel} 
(\mathcal{E}_{\bf k} \delta_{\alpha\beta} + e {\bf E} \cdot \mathcal{A}_{{\bf k},\alpha\beta}) 
\nonumber \\
&\equiv \mathcal{E}^{nm}({\bf k}_\perp)\delta_{\alpha\beta} +e {\bf E} \cdot \mathcal{A}_{\alpha\beta}^{nm}({\bf k}_\perp)
\end{align}
with $\alpha,\beta \in \{u,d\}$ and $n,m \in \mathbb{Z}$. 
In general, this Floquet-type eigenvalue equation is solved via numerical digonalization.

Since actual solutions are obtained numerically, it is not easy to figure out the precise dependence of $\mathcal{E}^{\rm WSL}({\bf k}_\perp)$ on ${\bf k}_\perp$.
It is, however, possible to infer a general structure of $\mathcal{E}^{\rm WSL}({\bf k}_\perp)$ based on the following two observations.
First, if $\mathcal{E}^{\rm WSL}({\bf k}_\perp)$ is a solution, $\mathcal{E}^{\rm WSL}({\bf k}_\perp)+neEa_\parallel$ with $n$ being any integer is also a solution.
This means that $\mathcal{E}^{\rm WSL}({\bf k}_\perp)$ has a ladder-like structure even in the non-Abelian case.
Second, in the Abelian limit, i.e., when $\mathcal{A}_{{\bf k},\alpha\beta}=\mathcal{A}_{\bf k} \delta_{\alpha\beta}$, $\mathcal{E}^{\rm WSL}({\bf k}_\perp)$ recovers the semiclassical solution in Eq.~\eqref{WSL_spectrum}.
To see this, let us rewrite the eigenvalue equation in Eq.~\eqref{Heff_eigenvalue_eq} in a differential equation form:
\begin{align}
\left[ {\cal E}_{\bf k}+eE\left( i\frac{\partial}{\partial k_\parallel}+{\cal A}_{{\bf k},\parallel}  \right) \right] \eta({\bf k})
={\cal E}^{\rm WSL}({\bf k}_\perp) \eta({\bf k}) ,
\end{align}
where ${\cal A}_{{\bf k},\parallel}$ is the parallel component of ${\cal A}_{\bf k}$ along the electric field.
The above equation can be solved formally by 
\begin{align}
\eta({\bf k}) = e^{
-\frac{i}{eE} \int_0^{k_\parallel} dk^\prime_\parallel 
\left[  
{\cal E}^{\rm WSL}({\bf k}_\perp)
-{\cal E}_{{\bf k}^\prime} 
-eE{\cal A}_{{\bf k}^\prime,\parallel} 
\right]
} ,
\label{Abelian_solution}
\end{align} 
where ${\bf k}=({\bf k}_\perp,k_\parallel)$ and ${\bf k}^\prime=({\bf k}_\perp,k^\prime_\parallel)$.
Equation~\eqref{Abelian_solution} is not yet a complete solution since the energy eigenvalue ${\cal E}^{\rm WSL}({\bf k}_\perp)$ is unknown.
${\cal E}^{\rm WSL}({\bf k}_\perp)$ is determined by imposing the periodic boundary condition:
$\eta({\bf k}_\perp, k_\parallel+2\pi/a_\parallel)=\eta({\bf k}_\perp, k_\parallel)$, which leads to the following quantization condition:
\begin{align}
\frac{1}{eE} \int_0^{2\pi/a_\parallel} dk^\prime_\parallel 
\left[  
{\cal E}^{\rm WSL}({\bf k}_\perp)
-{\cal E}_{{\bf k}^\prime} 
-eE{\cal A}_{{\bf k}^\prime,\parallel} 
\right]
= 2n\pi ,
\end{align}
which can be rearranged as the semiclassical solution in Eq.~\eqref{WSL_spectrum}.
Based on these two observations, we deduce a general form of $\mathcal{E}^{\rm WSL}({\bf k}_\perp)$ as follows:
\begin{align}
\mathcal{E}^{\rm WSL}_n({\bf k}_\perp)
& = \tilde{\mathcal{E}}({\bf k}_\perp) + \left( n + \frac{\tilde{\gamma}({\bf k}_\perp)}{2\pi} \right) eE a_\parallel , 
\label{Non_Abelian_WSL_spectrum}
\end{align}
where it is assumed that both $\tilde{\mathcal{E}}$ and $\tilde{\gamma}$ do not depend on $E$ at sufficiently weak $E$, i.e., in the adiabatic limit. 
This means actually that $\tilde{\mathcal{E}} = \bar{\mathcal{E}} = \frac{a_\parallel}{2\pi} \oint_C dk_\parallel \mathcal{E}_\mathbf{k}$.
Meanwhile, $\tilde{\gamma}$ reduces to $\gamma_{\rm Zak}$ only if the Berry connection/curvature becomes Abelian.
It is confirmed in Sec.~\ref{sec:3D_TI} that the exact eigenvalue solution of the requantized effective Hamiltonian is indeed well described by Eq.~\eqref{Non_Abelian_WSL_spectrum} (and is also entirely consistent with the full quantum theory presented in the following section).

%%%%%%%%%%%%%%%%%%%%%%%%%%%%%%%%%%%%%%%%%%%%%%%%%%%%%%%%%%%%%%%%%%%%%
%%%%%%%%%%%%%%%%%%%%%%%%%%%%%%%%%%%%%%%%%%%%%%%%%%%%%%%%%%%%%%%%%%%%%
\begin{figure*}[t]
\centering
\includegraphics[width=0.8\textwidth]
{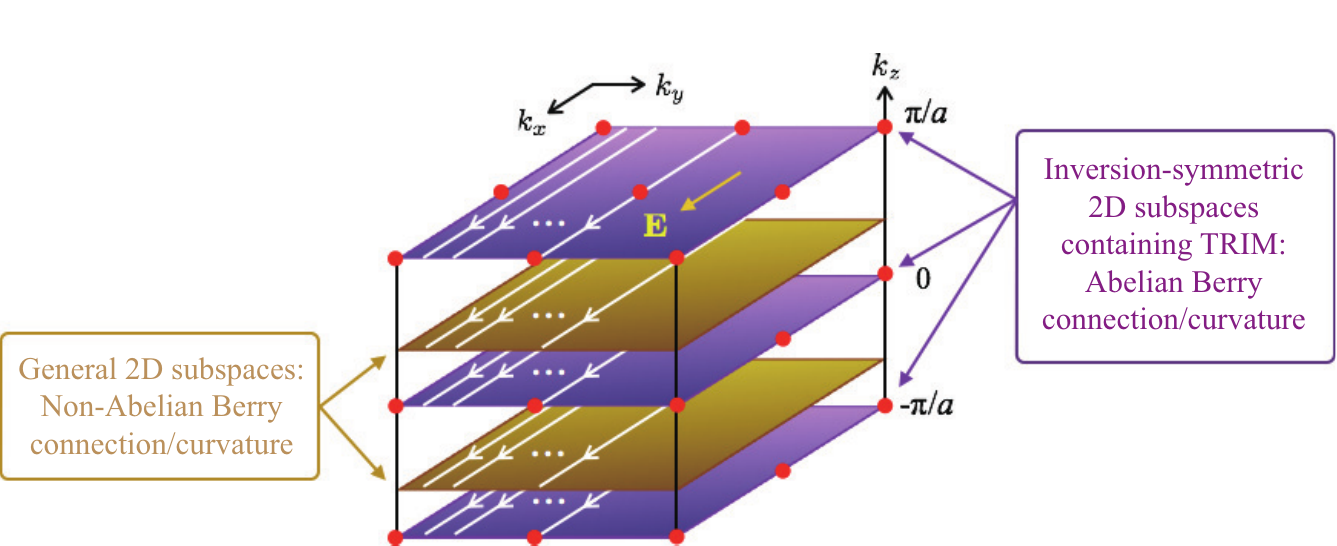} \\
\caption{
Schematic diagram illustrating how the 3D topological invariants are directly manifested in the winding numbers of the WSL within the inversion-symmetric 2D subspaces (purple-colored planes) containing TRIM (denoted by red circles), where the Berry connections/curvatures become Abelian. 
Here, the 2D subspaces are chosen to be parallel to the $k_x\mbox{--}k_y$ plane within the 3D BZ without loss of generality. 
According to Eq.~\eqref{3D_Z2_invariant}, the strong $\mathbb{Z}_2$ invariant is determined by the two $\mathbb{Z}_2$ invariants of the 2D subspaces containing one set of four TRIM at $k_z a=0$ and the other at $k_z a=\pi$. 
In general 2D subspaces (gold-colored planes), the Berry connections/curvatures are non-Abelian and thus the WSL does not have well-defined winding numbers.
} 
\label{Fig2_Non-Abelian_schematic}
\end{figure*}
%%%%%%%%%%%%%%%%%%%%%%%%%%%%%%%%%%%%%%%%%%%%%%%%%%%%%%%%%%%%%%%%%%%%%
%%%%%%%%%%%%%%%%%%%%%%%%%%%%%%%%%%%%%%%%%%%%%%%%%%%%%%%%%%%%%%%%%%%%%

So far, we have explained how to obtain the energy spectrum of the WSL eigenstates in the general case of the non-Abelian connection/curvature.  
With all the complications due to the non-Abelian structure, however, it is unclear at this stage how the energy spectrum of the WSL eigenstates can manifest the band topology in three dimensions.
Fortunately, in the presence of both inversion and time-reversal symmetries, the 3D band topology is characterized by the band topologies of the special 2D subspaces within the 3D BZ, where the Berry connections/curvatures become Abelian.
These special 2D subspaces are those with the inversion symmetry, where the inversion parity plays the role of a good quantum number guaranteeing the existence of a diagonalized basis.     
In other words, within these special 2D subspaces, the Berry connection satisfies the following Abelian condition,
\begin{align}
[\mathcal{A}_{\bf k}^i, \mathcal{A}_{\bf k}^j] = 0 , 
\label{AbelianCondition}
\end{align}
with $i$ and $j$ denoting two orthogonal directions in the 2D subspaces. 
In the cubic lattice, these 2D subspaces are the $k_i\mbox{--}k_j$ planes with $k_k a=0$ and $\pi$, where $(i,j,k) = (x,y,z)$, $(y,z,x)$, or $(z,x,y)$.
Under this condition, one can always find an appropriate set of bases, via which both $\mathcal{A}_{\bf k}^i$ and $\mathcal{A}_{\bf k}^j$ are diagonalized simultaneously.
Then, the WSL eigenenergy, $\mathcal{E}^{\rm WSL}_\sigma({\bf k}_\perp)$, for each component of the diagonalized basis, say, pseudospin $\sigma$, simply reduces to the Abelian version with a well-defined Chern number.
Consequently, the WSL has a well-defined winding number for each diagonalized pseudospin component in these 2D subspaces.

Now, with the time-reversal symmetry dictating that the total Chern number is zero, the three weak $\mathbb{Z}_2$ invariants are simply the 2D $\mathbb{Z}_2$ invariants of three 2D subspaces at $k_x=0$, $k_y=0$, and $k_z=0$, which are half the Chern number differences between diagonalized pseudospin components within the corresponding 2D subspaces. 
Similarly, according to Eq.~\eqref{3D_Z2_invariant}, the strong $\mathbb{Z}_2$ invariant is determined  by the two $\mathbb{Z}_2$ invariants of 2D subspaces, for example, at $k_z a=0$ and $\pi$.
As mentioned previously, the same strong $\mathbb{Z}_2$ invariant is obtained regardless of whether one chooses the $k_x a=0$ and $\pi$, the $k_y a=0$ and $\pi$ , or the $k_z a=0$ and $\pi$ planes.

In addition to the general argument for the Abelian condition above, one can explicitly show that the Abelian condition is precisely satisfied in the generic four-band model in Eq.~\eqref{Hamiltonian3DTI} for 3D TIs.
Actual calculations are a little bit messy, but the gist of why the Abelian condition is satisfied in this model can be revealed by examining the long-wavelength expansion of $\mathcal{A}_{\bf k}$ around TRIM.  
For convenience, let us concentrate on one of the TRIM at ${\bf k}=0$, where $\mathcal{A}_{\bf k}$ is expanded as
\begin{align}
\mathcal{A}_{\bf k} \simeq \xi_{\bf k} \sigma_z \hat{z} \times {\bf k} + \xi^\prime_{\bf k} \boldsymbol{\sigma} \times {\bf k},
\label{A_expansion}
\end{align}
where the concrete forms of the coefficients, $\xi_{\bf k}$ and $\xi^\prime_{\bf k}$, are not important for the current purpose.
With help of Eq.~\eqref{A_expansion}, it is now straightforward to show that the Abelian condition is satisfied for all three 2D subspaces of the $k_x\mbox{--}k_y$, $k_y\mbox{--}k_z$, and $k_z\mbox{--}k_x$ planes around ${\bf k}=0$. 
Similar arguments can be given for other TRIM.
As mentioned, it can be explicitly shown without any expansion that the Abelian condition is precisely satisfied for the entire subspaces of inversion-symmetric planes. 
%%%%%%%%%%%%%%%%%%% Reference to Appendix B %%%%%%%%%%%%%%%%%%%%%%%%%
See Appendix~\ref{Appen:Abelian_condition} for the proof of the Abelian condition in the generic four-band model.
%%%%%%%%%%%%%%%%%%%%%%%%%%%%%%%%%%%%%%%%%%%%%%%%%%%%%%%%%

To summarize, the 3D topological invariants are directly manifested in the winding numbers of the WSL within the inversion-symmetric 2D subspaces containing TRIM, where the Berry connections/curvatures become Abelian.  
In general 2D subspaces, the WSL does not have well-defined winding numbers.
Regardless of the existence of well-defined winding numbers, however, it is always possible to predict the energy spectrum of the WSL eigenstates precisely by solving the eigenvalue equation of the requantized effective Hamiltonian in Eq.~\eqref{Heff_eigenvalue_eq}.
This means that the requantized effective theory by itself can be applied to any situations including 2D/3D topological insulators even with time-reversal symmetry breaking terms.
See Fig.~\ref{Fig2_Non-Abelian_schematic} for a schematic diagram illustrating the discussions so far in this section.

In our theory, the existence of well-defined 3D topological invariants depends crucially on the fact that the the Berry connections/curvatures become Abelian and thus the winding numbers of the WSL are well defined within the inversion-symmetric 2D subspaces containing TRIM.
It is important to note that, similar to our theory, the logic behind {\it dimensional increase} for the derivation of 3D $\mathbb{Z}_2$ invariants is also based on the fact that the 2D topological invariants are well defined within the inversion-symmetric 2D subspaces containing TRIM.
Interestingly, the discrete representation of the strong $\mathbb{Z}_2$ invariant in Eq.~\eqref{3D_Z2_invariant} becomes equivalent to an integral representation formulated in terms of the Chern-Simons three form if both inversion and time-reversal symmetries are present~\cite{Wang10}.
Concretely, the integral of the Chern-Simons three form over the 3D BZ is given as follows~\cite{Qi08b}:
\begin{align}
\theta
= \frac{1}{8\pi} \int d^3 k~\epsilon^{ijk} {\rm Tr} \left[ \left( \mathcal{B}_{\bf k}^{ij} + \frac{2}{3} i \mathcal{A}_{\bf k}^i \cdot \mathcal{A}_{\bf k}^j \right) \cdot \mathcal{A}_{\bf k}^k \right], 
\label{CS3form}
\end{align}
where the non-Abelian Berry curvature is given by $\mathcal{B}_{\bf k}^{k} = \frac{1}{2} \epsilon^{ijk} \mathcal{B}_{\bf k}^{ij}$, where $\mathcal{B}_{\bf k}^{ij} = \partial_{k_i} \mathcal{A}_{\bf k}^j - \partial_{k_j} \mathcal{A}_{\bf k}^i - i [\mathcal{A}_{\bf k}^i, \mathcal{A}_{\bf k}^j]$ with $i,j,k \in \{x,y,z\}$.
In general, $\theta$ in Eq.~\eqref{CS3form} is not quantized in contrast to the Chern number.
Fortunately, however, in a close analogy with the fact that the one-dimensional Zak phase is quantized in the presence of the inversion symmetry~\cite{Zak89},
it can be proved that $\theta$ is quantized and related with the strong $\mathbb{Z}_2$ invariant $\nu_0$ in the presence of both inversion and time-reversal symmetries~\cite{Wang10}:
\begin{align}
\nu_0 = \frac{\theta}{\pi} \;\; {\rm (mod \;\; 2)} .
\label{connection_nu0_theta}
\end{align}
%It is important to note that the time-reversal symmetry is important in this proof since it guarantees the existence of a global basis with diagonalized (pseudo)spins in the TRIM-containing 2D subspaces, where the Berry connections are Abelian. 
In a sense, $\theta$ can be regarded as a 3D analog of the one-dimensional Zak phase. 
It is important to note that the presence of both inversion and time-reversal symmetries is crucial to guarantee the existence of a global basis of wave functions, which is necessary to prove Eq.~\eqref{connection_nu0_theta}.

%In the next section, we discuss the full quantum theory, which can be applied to general situations even beyond the adiabatic limit.
Finally, it is worthwhile to mention that the $\mathbb{Z}_2$ invariants can be also obtained via the {\it Wilson loop}, which is the non-Abelian generalization of the Zak phase factor~\cite{Grusdt14, Yu11}; 
\begin{align}
\hat{{\cal W}} = {\cal P} \exp{\left( i \oint_C d {\bf k}_{\parallel} \cdot {\cal A}_{\bf k} \right)} ,
\label{Wilson_loop}
\end{align}
where ${\cal P}$ is the path ordering operator, ${\cal A}_{\bf k}$ is the non-Abelian Berry connection, and $C$ denotes a closed path traced by ${\bf k}_\parallel$.
(Note that a different sign convention is used for the phase in Eq.~\eqref{Wilson_loop} compared to Refs.~\cite{Grusdt14, Yu11})
Concretely, the $\mathbb{Z}_2$ invariants are related with the winding numbers of the phase of the eigenvalues of the Wilson loop across the one-dimensional BZ of ${\bf k}_\perp$~\cite{Yu11}.
Considering that the phase of the eigenvalues of the Wilson loop is in turn related with the center position of the Wannier function, this method is highly reminiscent of ours, where the $\mathbb{Z}_2$ invariants are related with the winding numbers of the WSL.
Below, we explain exactly how these two methods are connected.

To this end, it is important to understand physically what the Wilson loop means.
The physical meaning of the Wilson loop was elucidated by Grusdt et al.~\cite{Grusdt14}, who have shown that the Wilson loop is nothing but the propagator describing the Bloch oscillation in the limit of strong electric field. 
To appreciate further what this means precisely, let us examine the propagator at general strengths of electric field (for a closed path $C$):
\begin{align}
\hat{\cal U}= {\cal P} \exp{\left[ i \oint_C d {\bf k}_\parallel \cdot \left( {\cal A}_{\bf k} +\frac{1}{eE} H_{\bf k} \right)  \right]} ,
\label{Propagator}
\end{align}  
where $H_{\bf k}$ is the Hamiltonian in the absence of electric field.
In the specific situation with two degenerate energy bands described in this section, $H_{\bf k}={\cal E}_{\bf k} \mathbb{I}_2$.
See Appendix C in Ref.~\cite{Grusdt14} for a detailed derivation of Eq.~\eqref{Propagator}.

Now, it is important to note that, while expressed in a different gauge (namely, the time-dependent vector potential gauge via the Peierls substitution), 
$\hat{\cal U}$ is actually equivalent to the propagator of our requantized effective Hamiltonian ${\cal H}_{\rm eff}$ in Eq.~\eqref{Heff} (expressed in the time-independent scalar potential gauge).
In this context, a rationale behind the usefulness of the Wilson loop is understood as follow.
Since the main information on band topology is embedded in ${\cal A}_{\bf k}$, one may ignore $H_{\bf k}$ from the argument of the exponential in Eq.~\eqref{Propagator} if one is only interested in the characterization of the band topology, not the detailed coherent dynamics of the Bloch oscillation. 
Roughly, diagonalizing the Wilson loop is equivalent to diagonalizing ${\cal H}_{\rm eff}$ while ignoring the band-dispersion part.
Therefore, the winding numbers of the phase of the eigenvalues of the Wilson loop should carry essentially the same topological information as those of the WSL studied in this work.

Rigorously, however, $H_{\bf k}$ can be ignored only in the limit of strong electric field, i.e., $\hat{\cal U}_{E=\infty}=\hat{\cal W}$, where the adiabatic condition, which is necessary for the very validity of the Berry connection/curvature, is completely violated.
Therefore, the winding numbers of the phase of the eigenvalues of the Wilson loop may not be physically observable due to the contradiction between the adiabatic condition and the strong electric-field limit. 
An improvement of the Wilson-loop scheme is to use directly the coherent dynamics of the Bloch oscillation, which is governed by $\hat{\cal U}_{E \neq \infty}$.
As mentioned in Sec.~\ref{sec:Introduction}, an interferometric method combining the coherent Bloch oscillation with the Ramsey interferometry was recently proposed to measure the $\mathbb{Z}_2$ invariants in optical lattices~\cite{Grusdt14}.
It is emphasized, however, that our method, where the winding numbers of the WSL are observed in spectroscopic measurements, can be applied to condensed matter systems, where the phase coherence is not guaranteed.

%%%%%%%%%%%%%%%
%% Full quantum theory %%
%%%%%%%%%%%%%%%
\subsection{Full quantum theory}
\label{sec:Full_quantum_theory}

The semiclassical theory and the subsequent requantized effective theory are valid under the condition that the Bloch oscillation energy $\hbar\Omega=eEa_\parallel$ is sufficiently smaller than the interband energy difference. 
This condition is nothing but the adiabatic condition for the validity of the Berry phase so that electrons can remain in a given band during the Bloch oscillation without making transitions to other bands.
In other words, the electric field should be sufficiently weak so that the non-linear effects are negligible. 
On the other hand, the Bloch oscillation should be sufficiently faster than the electron-impurity scattering rate so that electrons can complete a full cycle of the Bloch oscillation before scattered off. 
In what follows, we assess if these two conditions can be met simultaneously in the full quantum theory.

The energy spectrum of the WSL eigenstates is revealed as a series of sharp peaks in the density of states (DOS), which can be obtained as the imaginary part of the retarded Green's function, ${\rm Im} G^r_{\bf k}(\omega)$. 
There are two gauge choices for the implementation of an electric field; (i) the time-independent scalar potential gauge with $\phi=-{\bf E}\cdot{\bf x}$ and (ii) the time-dependent vector potential gauge with $\mathbf{A}=-c\mathbf{E}t$.
In this work, we choose the time-dependent vector potential gauge, where the spatial translation symmetry is formally present. 
In the formal presence of the spatial translation symmetry, the momentum parallel to the electric field, $k_\parallel$, is a conserved quantity. 
However, the DOS becomes gauge-invariant and physically meaningful only if it is integrated over $k_\parallel$. 
The physically meaningful semi-local DOS is obtained by integrating out ${\rm Im} G^r_{\bf k}(\omega)$ with respect to $k_\parallel$.
The Green's function in the frequency domain, $G^r_{\bf k}(\omega)$, is obtained from its counterpart in the time domain, $G^r_{\bf k}(t,t^\prime)$, via the Fourier transformation.
In the time-dependent vector potential gauge, $G^r_{\bf k}(t,t^\prime)$ is computed via the Peierls shift,  $\hbar\mathbf{k} \rightarrow \hbar\mathbf{k} - e\mathbf{E}t$, of the Hamiltonian $H({\bf k})$, i.e., $H(\mathbf{k}) \rightarrow H(\mathbf{k} - e\mathbf{E}t/\hbar)$. 
It is interesting to note that the Peierls shift plays a role of incorporating one of the two semiclassical equations of motion mentioned previously in Sec.~\ref{sec:Semiclassical_theory}.

Below, we explain how to compute the (retarded) Green's function of the time-dependent Hamiltonian via the Floquet Green's function formalism~\cite{Tsuji08,Lee14}.  
In the Floquet Green's function formalism, the semi-local DOS is computed as follows:
\begin{align}
\rho_\alpha ({\bf k}_\perp, \omega+n\tilde{\Omega})
= - \frac{1}{\pi} \mathrm{Im} \sum_{k_\parallel} (G_{\bf k}^r)_{\alpha\alpha}^{nn}(\omega) ,
\label{SemilocalDOS}
\end{align}
where the Floquet Green's function is given by
\begin{align}
(G_{\bf k}^r)_{\alpha\beta}^{nm}(\omega) = \int dt \int dt' e^{i(\omega + n\tilde{\Omega})t} e^{-i(\omega + m\tilde{\Omega})t'} (G_{\bf k})_{\alpha\beta}^r(t,t')
\label{FloquetGreensFuncDef}
\end{align}
with $-\tilde{\Omega}/2 < \omega \leq \tilde{\Omega}/2$.
Here, the Floquet frequency $\tilde{\Omega}$ is the natural frequency of the Peierls-shifted Hamiltonian, $H(\mathbf{k} - e\mathbf{E}t/\hbar)$, which depends on the specific structure of the Hamiltonian and is not necessarily the same as $\Omega$.
Usually, $\rho_\alpha$ is summed over $\alpha$ since we are interested in the semi-local DOS contributed by all generalized orbitals including both spin and orbital degrees of freedom.
However, in 2D TIs, where perpendicular spin components are conserved, $\rho_\alpha$ for each spin component can be meaningful.
Furthermore, in the special case of the Kane-Mele model defined on the honeycomb lattice, where the sublattice index is spatially distinguishable, $\rho_\alpha$ for each spin and sublattice index can be independently meaningful.

The retarded Green's function in the time domain, $(G^r_{\bf k})_{\alpha\beta}(t,t')$, is obtained by solving the following equation:
\begin{align}
%\frac{\partial}{\partial t} G^r_{\mathbf{k},\alpha\beta}(t,t') 
%= &-\frac{i}{\hbar} \delta_{\alpha\beta} \delta(t-t') 
%\nonumber\\
%&- \frac{i}{\hbar} \sum_\gamma H_{\alpha\gamma}(\mathbf{k} - e\mathbf{E}t/\hbar ) G^r_{\mathbf{k},\gamma\beta}(t,t'),
\frac{\partial}{\partial t} (G^r_{\bf k})_{\alpha\beta}(t,t') 
= &-\frac{i}{\hbar} \delta_{\alpha\beta} \delta(t-t') 
\nonumber\\
&- \frac{i}{\hbar} \sum_\gamma H_{\alpha\gamma}(\mathbf{k} - e\mathbf{E}t/\hbar ) (G^r_{\bf k})_{\gamma\beta}(t,t'),
\label{GreensFuncEq}
\end{align}
where $H_{\alpha\beta}({\bf k}- e\mathbf{E}t/\hbar)$ is the matrix element of the Peierls-shifted Hamiltonian between generalized orbital $\alpha$ and $\beta$ at momentum ${\bf k}$. 
%Note that, here, $\alpha$, $\beta$, and $\gamma$ indicate the generalized orbital indices including both actual spin and orbital degrees of freedom. 
Moving to the frequency domain via the Fourier transformation, Eq.~\eqref{GreensFuncEq} can be written in the Floquet matrix form:
\begin{align}
&\sum_{l,\gamma} 
\left\{ \left[\hbar(\omega + n\tilde{\Omega}) + i\eta\right] \delta_{\alpha\gamma}\delta_{nl}  - H^{nl}_{\alpha\gamma}(\mathbf{k}) \right\} 
(G^r_{\bf k})_{\gamma\beta}^{lm}(\omega)
\nonumber \\
&\;\;\;\;\;= \delta_{\alpha\beta} \delta_{nm},
\label{FloquetGreensFuncEq}
\end{align}
where $H^{nm}_{\alpha\beta}({\bf k}) = \frac{1}{T} \int_0^T dt e^{i(n-m)\tilde{\Omega} t} H_{\alpha\beta}( \mathbf{k} - e\mathbf{E}t/\hbar )$ with $T = 2\pi/\tilde{\Omega}$.
%%%%%%%%%%%%%%%%%%%%%% Reference to Appendix C %%%%%%%%%%%%%%%%%%%%%%%%%%%%
See Appendix~\ref{Appen:FloquetGreensFunc} for details on how to compute $(G^r_{\bf k})_{\alpha\beta}^{nm}(\omega)$ from Eq.~\eqref{FloquetGreensFuncEq} for various TI models. %(studied in Sec.~\ref{sec:Results}). 
%%%%%%%%%%%%%%%%%%%%%%%%%%%%%%%%%%%%%%%%%%%%%%%%%%%%%%%%%%%%%%%

The Floquet Green's function formalism provides a natural platform to study the effects of electron-impurity scattering~\cite{Lee14}.
To this end, we take a simple model Hamiltonian for the non-magnetic on-site electron-impurity interaction, 
\begin{align}
H_\mathrm{imp} = V \sum_{i,\alpha} n_{i\alpha} n_{{\rm imp}, i},
\end{align} 
where $V$ is the electron-impurity interaction strength, and $n_{i\alpha}$ and $n_{{\rm imp}, i}$ are the electron and impurity number operators at the $i$-th lattice site, respectively. 
The full Green's function, $\mathbb{G}^r_{\bf k}$, is obtained by solving the Dyson equation 
\begin{align}
(\mathbb{G}^{r -1}_{\bf k})_{\alpha\beta}^{nm} = (G^{r -1}_{{\bf k}})_{\alpha\beta}^{nm} - (\Sigma^r)_{\alpha\alpha}^{nn}\delta_{\alpha\beta}\delta_{nm}, 
\end{align}
where $(G^{r-1}_{\bf k})_{\alpha\beta}^{nm}$ is the inverse of the non-interacting Green's function computed in Eq.~\eqref{FloquetGreensFuncEq} and $\Sigma^r$ is the self-energy due to the electron-impurity interaction.
In this work, the impurity self-energy is computed via the self-consistent Born approximation (SCBA): 
\begin{align}
(\Sigma^r)_{\alpha\alpha}^{nn}(\omega) =  V_\textrm{imp}^2 \sum_\mathbf{k} (\mathbb{G}^r_{\bf k})_{\alpha\alpha}^{nn}(\omega), 
\label{SCBA_Self-energy}
\end{align}
where $V_\textrm{imp} \equiv \sqrt{\bar{n}_\textrm{imp}}~V$ with $\bar{n}_\textrm{imp}$ being the average impurity number per site.
It is important to note that, in principle, this formalism can be also applied to strongly correlated topological insulators, once the accurate self-energy is obtained for a strong electron-electron interaction.

In what follows, it is shown that the results of the full quantum theory are entirely consistent with those of the semiclassical theory in the Abelian case and the requantized effective theory in the general non-Abelian case with all being robust against interband interference as well as non-magnetic impurity scattering.

%%%%%%%%%%%
%%% Results %%%
%%%%%%%%%%%
\section{Results}
\label{sec:Results}

%%%%%%%%
%% 2D TI %%
%%%%%%%%

\subsection{2D TI}
\label{sec:2D_TI}

%%%%%%%%%
% BHZ model %
%%%%%%%%%

\subsubsection{Bernevig-Hughes-Zhang (BHZ) model}

We study the BHZ model~\cite{Bernevig06} as a first example of the 2D TI model with conserved perpendicular spin components. 
Imposed by the symmetry of the underlying microscopic structure around the $\Gamma$ point, the Hamiltonian for four low-lying states $(E_1\!\uparrow,H_1\!\uparrow,E_1\!\downarrow,H_1\!\downarrow)$ can be written as the generic form in Eq.~\eqref{Hamiltonian2DTI}, which is repeated here for convenience: 
$H = \sum_{\mathbf{k}, \sigma=\uparrow,\downarrow} \psi_{\mathbf{k}\sigma}^\dag H_\sigma(\mathbf{k}) \psi_{\mathbf{k}\sigma}$ with 
$H_\downarrow(\mathbf{k}) = H_\uparrow^*(-\mathbf{k})$ and 
\begin{align}
H_\uparrow(\mathbf{k})
& = \epsilon_\mathbf{k} \mathbb{I}_2 + \mathbf{d}_\mathbf{k} \cdot \boldsymbol{\sigma} 
= \left(
\begin{array}{cc}
\epsilon_\mathbf{k} + d_{\mathbf{k},z}
& d_{\mathbf{k},-} \\
d_{\mathbf{k},+}
& \epsilon_\mathbf{k} - d_{\mathbf{k},z}
\end{array}
\right),
\label{BHZ_Hamiltonian}
\end{align}
where $\psi_{\mathbf{k}\sigma}^\dag = (c_{\mathbf{k}E_1\sigma}^\dag, c_{\mathbf{k}H_1\sigma}^\dag)$ with $c_{\mathbf{k}\alpha\sigma}^\dag$ being the electron creation operator with momentum $\mathbf{k}$ and spin $\sigma$ $(=\uparrow,\downarrow)$ on orbital $\alpha$ $(=E_1,H_1)$.  
Around the $\Gamma$ point, $\epsilon_{\bf k}$ and ${\bf d}_{\bf k}$ can be expanded as  
$\epsilon_\mathbf{k} \simeq C + D (k_x^2 + k_y^2)$ and $\mathbf{d}_\mathbf{k} \simeq (A k_x, - A k_y, M + B (k_x^2 + k_y^2))$, respectively.

%%%%%%%%%%%%%%%%%%%%%%%%%%%%%%%%%%%%%%%%%%%%%%%%%%%%%%%%%%%%%%%%%%%%%%
%%%%%%%%%%%%%%%%%%%%%%%%%%%%%%%%%%%%%%%%%%%%%%%%%%%%%%%%%%%%%%%%%%%%%%
\begin{figure}[t]
\centering
\includegraphics[width=0.45\textwidth]
{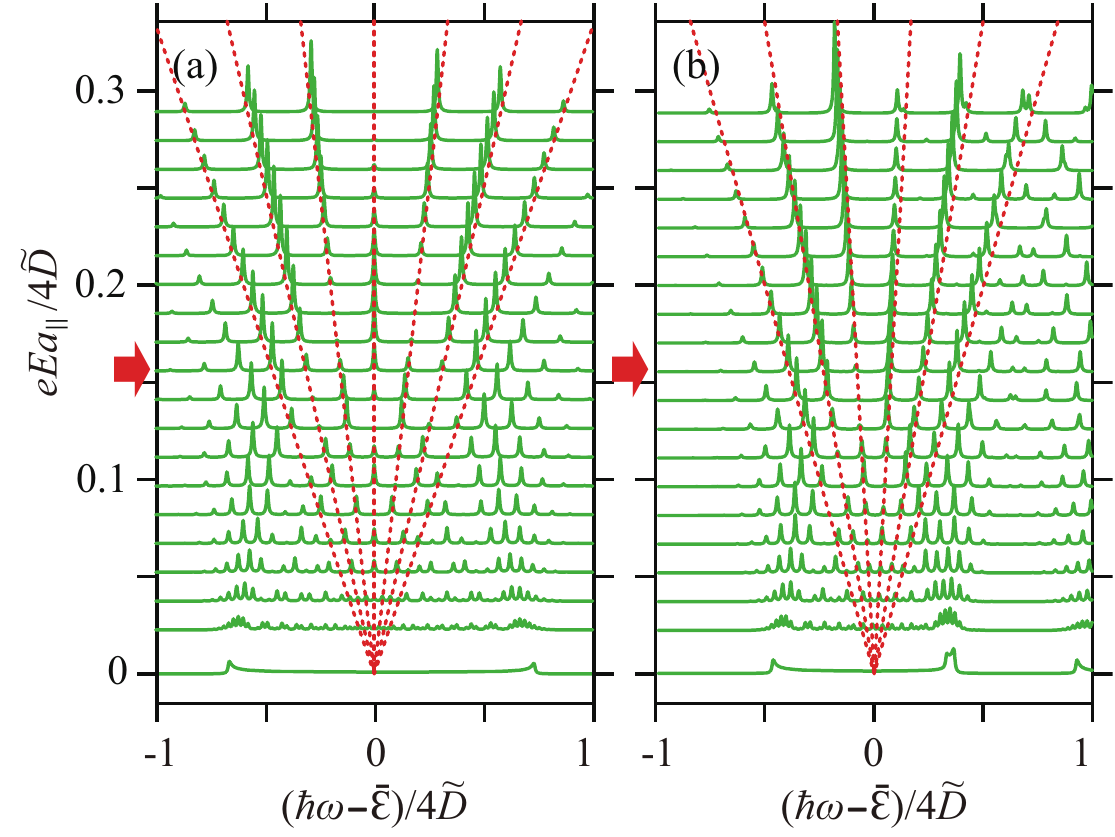} \\
\caption{
Evolution of the semi-local DOS at (a) $k_\perp a =\pm\pi$ and (b) $k_\perp a =0$ as a function of electric field. 
Note that fan-shaped series of the WSL branches emerge from the center of the valence band, being entirely consistent with the semiclassical theory in Sec.~\ref{sec:Semiclassical_theory}. 
Here, we consider a topologically non-trivial phase in the BHZ model with model parameters such as $\tilde{A}/4\tilde{D} = 0.6$, $\tilde{B}/4\tilde{D} = 0.6$, $C/4\tilde{D}=0$, and $M/4\tilde{D} = -0.3$. 
The red dashed lines are the semiclassical guide lines for the WSL branches obtained from Eq.~\eqref{WSL_spectrum}.
}
\label{Fig3_WSL_branches}
\end{figure}
%%%%%%%%%%%%%%%%%%%%%%%%%%%%%%%%%%%%%%%%%%%%%%%%%%%%%%%%%%%%%%%%%%%%%%
%%%%%%%%%%%%%%%%%%%%%%%%%%%%%%%%%%%%%%%%%%%%%%%%%%%%%%%%%%%%%%%%%%%%%%

The low-energy Hamiltonian can be promoted to a tight-binding Hamiltonian via the minimal lattice regularization, which replaces $k$ by $\sin{(k a)}/a$ and $k^2$ by $2[1-\cos{(k a)}]/a^2$.
Specifically, after the minimal lattice regularization, we set
$\epsilon_\mathbf{k} = C + 2\tilde{D} [ 2 - \cos(k_x a) - \cos(k_y a) ]$,
$d_{\mathbf{k},\pm} = \tilde{A} [\sin(k_x a) \mp i \sin(k_y a)]$, 
$d_{\mathbf{k},z} = M + 2\tilde{B} [2 - \cos(k_x a) - \cos(k_y a)]$, where we define $\tilde{A} = A / a$, $\tilde{B} = B / a^2$, and $\tilde{D} = D / a^2$, which all have the same physical unit as $M$, i.e., energy. 
Here, we set the electric field to be aligned along the principal direction of the square lattice so that $a_\parallel = a_\perp = a$.
As mentioned previously, the band topology becomes non-trivial if $M/B <0$ and trivial otherwise.

%%%%%%%%%%%%%%%%%%%%%%%%%%%%%%%%%%%%%%%%%%%%%%%%%%%%%%%%%%%%%%%%%%%%%%
%%%%%%%%%%%%%%%%%%%%%%%%%%%%%%%%%%%%%%%%%%%%%%%%%%%%%%%%%%%%%%%%%%%%%%
\begin{figure}[t]
\centering
\includegraphics[width=0.45\textwidth]
{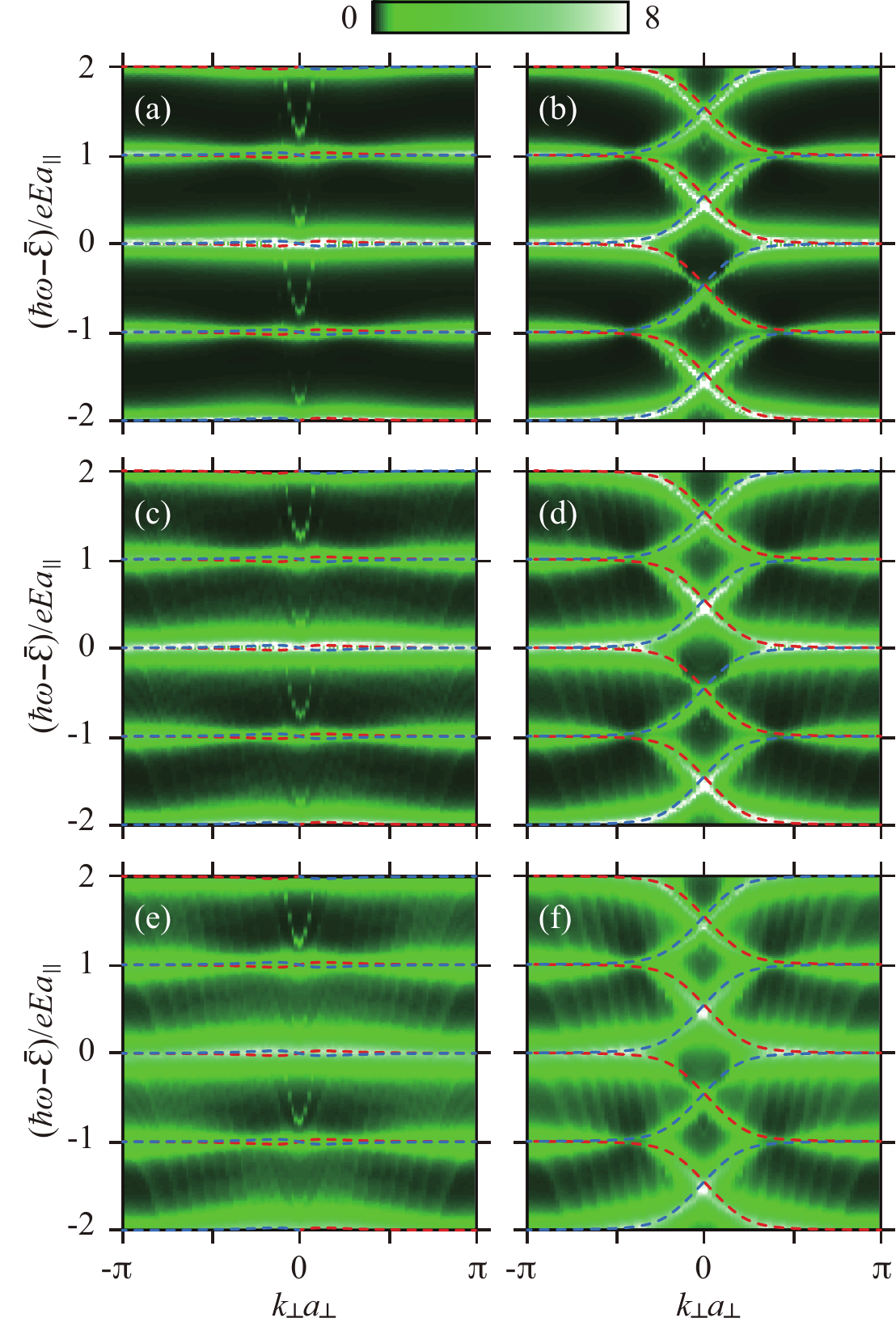} \\
\caption{
Semi-local DOS as a function of $k_\perp$ in the BHZ model. %showing that the 2D topological invariant is directly manifested in the winding number of the WSL.
Here, the semi-local DOS is summed over both spin and orbital degrees of freedom.
The electric field is applied along the principal direction of the square lattice with magnitude $eEa_\parallel/4\tilde{D} = 0.16$, which corresponds to the situations indicated  by the red arrows in Fig.~\ref{Fig3_WSL_branches}. 
Panels~(a), (c), and (e) denote when the band topology is trivial with a choice of $M/4\tilde{D} = 0.3$, while panels~(b), (d), and (f) denote when it is non-trivial with $M/4\tilde{D} = -0.3$. 
Here, $\tilde{A}/4\tilde{D}$, $\tilde{B}/4\tilde{D}$, and $C/4\tilde{D}$ are chosen as the same as those in Fig.~\ref{Fig3_WSL_branches}. 
The electron-impurity interaction strength is changed so that $V_\mathrm{imp} / 4\tilde{D} =$ 0, 0.05, and 0.08 in the top [(a) and (b)], middle [(c) and (d)], and bottom [(e) and (f)] panels, respectively.
The red and blue dashed lines denote the guide lines obtained in the semiclassical theory via Eq.~\eqref{WSL_spectrum} for spin up and down, respectively.
As predicted, the Kramers doublets exchange partners between $k_\perp a_\perp=0$ and $\pm\pi$ in the topologically non-trivial phase, while not in the trivial phase.
}
\label{Fig4_BHZ}
\end{figure}
%%%%%%%%%%%%%%%%%%%%%%%%%%%%%%%%%%%%%%%%%%%%%%%%%%%%%%%%%%%%%%%%%%%%%%
%%%%%%%%%%%%%%%%%%%%%%%%%%%%%%%%%%%%%%%%%%%%%%%%%%%%%%%%%%%%%%%%%%%%%%

Figure~\ref{Fig3_WSL_branches} shows the evolution of the semi-local DOS as a function of electric field, which exhibits fan-shaped series of the WSL branches emerging from the center of the valence band.
As one can see, there is excellent agreement between the results of the semiclassical theory via Eq.~\eqref{WSL_spectrum} for the Abelian Berry connection/curvature and those of the full quantum theory via Eq.~\eqref{SemilocalDOS}. 
It is important to note that, in addition to the main WSL branches emerging from the center, there are other WSL-like branches emerging from the band edges, which is known as the Franz-Keldysh effect~\cite{Schmidt94}.
Similarly, the conduction band (not shown in the figure) generates its own WSL eigenstate branches, which interfere with the WSL eigenstate branches emerging from the valence band at sufficiently strong electric fields. 
Fortunately, despite all these complicated interferences, the main WSL eigenstate branches emerging from the center of the valence band can be clearly identified at an appropriate window of electric field, say, $eEa_\parallel / 4\tilde{D} \simeq 0.16$, which is indicated by the red arrows in Fig.~\ref{Fig3_WSL_branches}.

Figure~\ref{Fig4_BHZ} shows the semi-local DOS at $eEa_\parallel / 4\tilde{D} = 0.16$ as a function of $k_\perp$, which confirms that the 2D topological invariant is directly manifested in the winding number of the WSL, precisely as predicted by the semiclassical theory in Eq.~\eqref{WSL_spectrum}.
Specifically, two separate sets of the spin-dependent WSL branches wind non-trivially and oppositely in the topologically non-trivial phase (right panels) accompanied by an exchange of the Kramers-doublet partners between $k_\perp a_\perp=0$ and $\pm\pi$. 
Meanwhile, there is no winding of the WSL in the trivial phase (left panels).

Now, to test the robustness of the band topology against non-magnetic impurity scattering, we investigate how the semi-local DOS changes as a function of electron-impurity interaction strength $V_{\rm imp}$. 
Note that, here, the effects of non-magnetic impurity scattering are taken into account within the SCBA via Eq.~\eqref{SCBA_Self-energy}. 
As one can see from Fig.~\ref{Fig4_BHZ}~(c)\mbox{--}(f), the winding number of the WSL can be clearly identified, unless the electron-impurity interaction strength becomes too strong to become comparable to the Bloch oscillation energy.

%%%%%%%%
% KM model %
%%%%%%%%
\subsubsection{Kane-Mele (KM) model}

%%%%%%%%%%%%%%%%%%%%%%%%%%%%%%%%%%%%%%%%%%%%%%%%%%%%%%%%%%%%%%%%%%%%%%
%%%%%%%%%%%%%%%%%%%%%%%%%%%%%%%%%%%%%%%%%%%%%%%%%%%%%%%%%%%%%%%%%%%%%%
\begin{figure}[t]
\centering
\includegraphics[width=0.45\textwidth]
{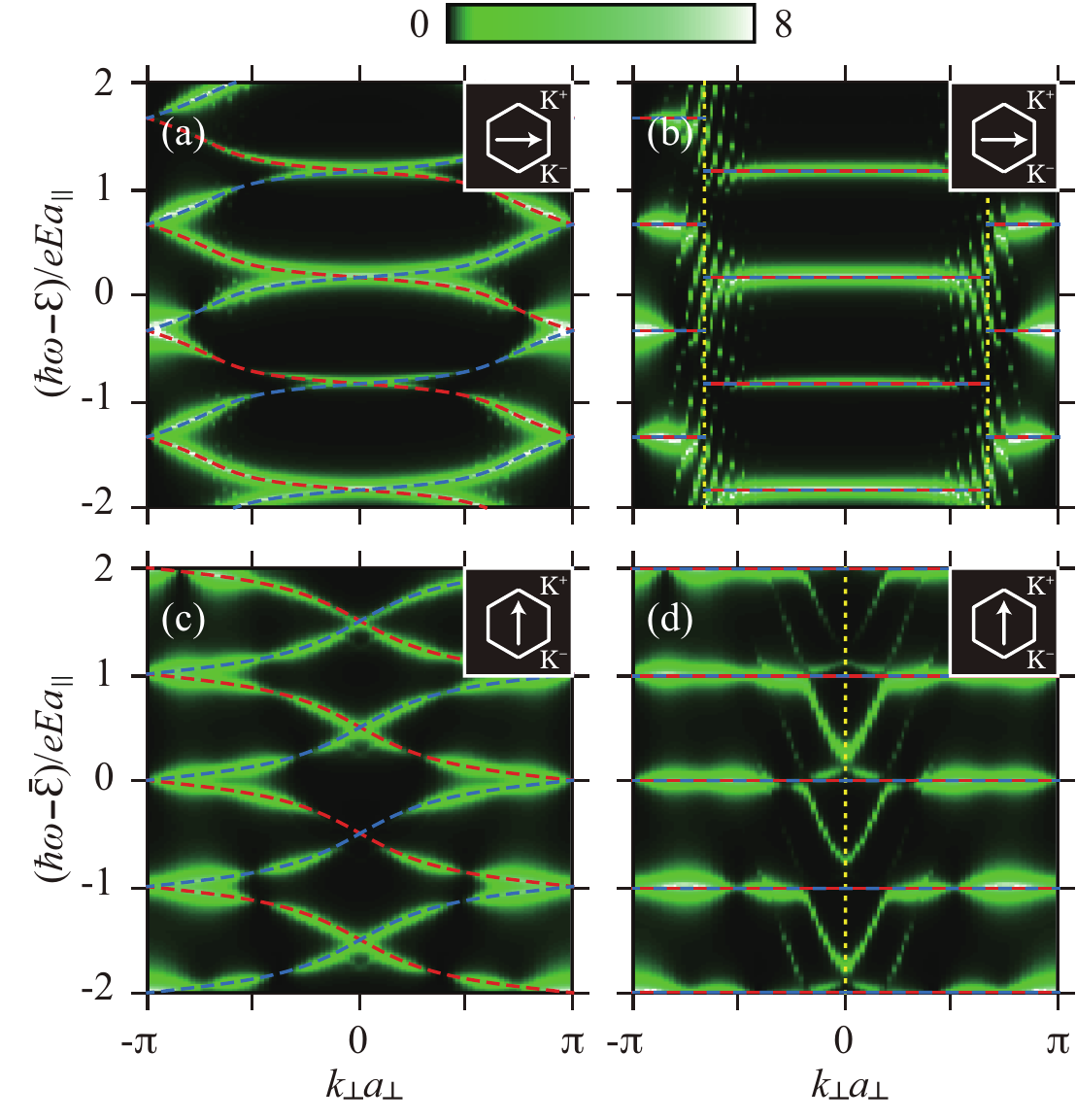} \\
\caption{
Semi-local DOS as a function of $k_\perp$ in the KM model. %showing that the 2D topological invariant is directly manifested in the winding number of the WSL.
Here, the semi-local DOS is shown only for a given sublattice of the honeycomb lattice, while summed over the spin degree of freedom.
The electric field with magnitude $eEa_\parallel/\tilde{t} = 0.16$ is applied along the armchair direction in panels (a) and (b), while along the zigzag direction in panels (c) and (d). 
See insets to see how the electric field is aligned in the BZ.
Panels (a) and (c) correspond to a topologically non-trivial phase with $\lambda_\mathrm{SO}/\tilde{t} = 0.1$, while panels (b) and (d) correspond to ordinary graphene with $\lambda_\mathrm{SO} = 0$. 
The red and blue dashed lines denote the guide lines obtained in the semiclassical theory via Eq.~\eqref{WSL_spectrum} for spin up and down, respectively.
The yellow dotted lines indicate when the semiclassical trajectory passes through the Dirac points, i.e., the monopole singularities.
}
\label{Fig5_KM}
\end{figure}
%%%%%%%%%%%%%%%%%%%%%%%%%%%%%%%%%%%%%%%%%%%%%%%%%%%%%%%%%%%%%%%%%%%%%%
%%%%%%%%%%%%%%%%%%%%%%%%%%%%%%%%%%%%%%%%%%%%%%%%%%%%%%%%%%%%%%%%%%%%%%

Next, we study the KM model~\cite{Kane05a, Kane05b}, whose Hamiltonian is defined on the honeycomb lattice as follows:
\begin{align}
H = - \tilde{t} \sum_{\langle i,j \rangle} \psi_{i}^\dag \psi_{j}
+i\lambda_\mathrm{SO} \sum_{\langle\langle i,j \rangle\rangle} \psi_{i}^\dag \nu_{ij} \sigma_z \psi_{j},
\label{KM_Hamiltonian}
\end{align}
where $\psi_{i}^\dagger = (c_{i \uparrow}^\dagger, c_{i \downarrow}^\dagger)$, $\tilde{t}$ is the hopping constant, and $\nu_{ij}=\pm 1$, depending on if the electron takes the left or right turn to get to the next-nearest neighbor.

After the Fourier transformation, Eq.~\eqref{KM_Hamiltonian} reduces to the generic 2D TI Hamiltonian in Eq.~\eqref{Hamiltonian2DTI} with 
$\epsilon_\mathbf{k} = 0$,
$d_{\mathbf{k}, \pm} =  - \tilde{t} ( e^{\mp i\mathbf{k} \cdot \mathbf{c}_1} + e^{\mp i\mathbf{k} \cdot \mathbf{c}_2} + e^{\mp i\mathbf{k} \cdot \mathbf{c}_3} )$, and 
$d_{\mathbf{k}, z} = 2\lambda_\mathrm{SO} [\sin(\mathbf{k} \cdot \mathbf{a}_1) - \sin(\mathbf{k} \cdot \mathbf{a}_2) - \sin(\mathbf{k} \cdot (\mathbf{a}_1 - \mathbf{a}_2))]$,
where $\mathbf{c}_1 = a (1/2, \sqrt{3}/2)$, $\mathbf{c}_2 = a (1/2, -\sqrt{3}/2)$, $\mathbf{c}_3 = a (-1, 0)$, $\mathbf{a}_1 = a (3/2, \sqrt{3}/2)$, and $\mathbf{a}_2 = a (3/2, -\sqrt{3}/2)$. 
As mentioned previously, the band topology can be determined by examining the low-energy behaviors of $d_{{\bf k},z}$ around the Dirac points $\mathbf{K}^\pm = (\frac{2\pi}{3a}, \pm \frac{2\pi}{3\sqrt{3}a})$;
$d_{{\bf k},z} \simeq \mp 3\sqrt{3} \lambda_\mathrm{SO} \pm \frac{9}{4}\sqrt{3} \lambda_\mathrm{SO} [(q_x a)^2 + (q_y a)^2]$ at ${\bf k}={\bf K}^\pm+{\bf q}$. 
Low-energy behaviors of $d_{{\bf k},z}$ near both $\mathbf{K}^{\pm}$ satisfy the non-triviality condition if $\lambda_{\rm SO} \neq 0$, which means that the whole valence band becomes topologically non-trivial under this condition.

Figure~\ref{Fig5_KM} shows the comparison between the semi-local DOS of a topologically non-trivial phase (left panels) with $\lambda_{\rm SO} \neq 0$ and ordinary graphene (right panels) with $\lambda_{\rm SO}=0$, which again confirms that the 2D topological invariant is directly manifested in the winding number of the WSL. 
It is interesting to note that graphene does not have a well-defined value for the 2D topological invariant since the monopole singularities at $\mathbf{K}^{\pm}$ are located right within the 2D BZ.  
Figure~\ref{Fig5_KM}~(b) and (d) show that the WSL energy spectrum behaves irregularly, when the semiclassical trajectory at a given $k_\perp$ passes through the monopole singularities at $\mathbf{K}^{\pm}$, making the Zak phase discontinuous~\cite{Delplace11}.
In some sense, graphene can be regarded as being topologically critical, neither being topologically trivial nor non-trivial.

Now, it is important to check that the winding number of the WSL does not depend on the electric-field direction, while the detailed $k_\perp$-dependence of the WSL energy spectrum may.
To this end, in Fig.~\ref{Fig5_KM}~(a) and (b), the electric field is applied along the armchair direction with $a_\parallel = 3a/2$ and $a_\perp = \sqrt{3}a$, while, in Fig.~\ref{Fig5_KM}~(c) and (d), applied along the zigzag direction with $a_\parallel =\sqrt{3}a/2$ and $a_\perp=3a$. 
As one can see, the winding number of the WSL does not depend on the electric-field direction, being consistent with the fact that the winding number of the WSL is a topological quantity.

%%%%%%%%
%% 3D TI %%
%%%%%%%%
\subsection{3D TI}
\label{sec:3D_TI}

Finally, we study the 3D TI model describing strong 3D TIs occurring in BiSe-family materials~\cite{Hasan10,Qi11}. 
Around the $\Gamma$ point, the Hamiltonian for four low-lying states $(P1_z^{+}\!\uparrow,P2_z^{-}\!\uparrow,P1_z^{+}\!\downarrow,P2_z^{-}\!\downarrow)$ can be written as the generic form in Eq.~\eqref{Hamiltonian3DTI}, which is repeated here for convenience:
$H = \sum_\mathbf{k} \psi_\mathbf{k}^\dag H(\mathbf{k}) \psi_\mathbf{k}$ with
\begin{align}
& H(\mathbf{k})
= \epsilon_\mathbf{k} \mathbb{I}_4 + \mathbf{d}_\mathbf{k} \cdot \boldsymbol{\Gamma} 
\nonumber \\
&= \left(
\begin{array}{cccc}
\epsilon_\mathbf{k} - d_{\mathbf{k},3} & d_{\mathbf{k},4} & 0 & d_{\mathbf{k},-} \\
d_{\mathbf{k},4} & \epsilon_\mathbf{k} + d_{\mathbf{k},3} & d_{\mathbf{k},-} & 0 \\
0 & d_{\mathbf{k},+} & \epsilon_\mathbf{k} - d_{\mathbf{k},3} & - d_{\mathbf{k},4} \\
d_{\mathbf{k},+} & 0 & - d_{\mathbf{k},4} & \epsilon_\mathbf{k} + d_{\mathbf{k},3}
\end{array}
\right),
\label{BiSe_Hamiltonian}
\end{align}
where $\psi_{\bf k}^\dag = (c_{{\bf k}\alpha_1}^\dag, c_{{\bf k}\alpha_2}^\dag, c_{{\bf k}\alpha_3}^\dag, c_{{\bf k}\alpha_4}^\dag)$ with $c_{{\bf k}\alpha}^\dag$ being the electron creation operator with momentum $\mathbf{k}$ on generalized orbital $\alpha=(P1_z^{+}\!\uparrow, P2_z^{-}\!\uparrow, P1_z^{+}\!\downarrow, P2_z^{-}\!\downarrow)$. 
As mentioned before, ${\bf d}_{\bf k}$ can be expanded around the $\Gamma$ point as 
$d_{{\bf k},\pm}=d_{{\bf k},1} \pm i d_{{\bf k},2} \simeq A_1 (k_x \pm i k_y)$,
$d_{{\bf k},3} \simeq M+B_1(k_x^2+k_y^2)+B_2 k_z^2$,
$d_{{\bf k},4} \simeq A_2 k_z$, and
$d_{{\bf k},5} \simeq 0$.
Also, $\epsilon_{\bf k}$ can be expanded similarly as $\epsilon_\mathbf{k} \simeq C + D_1 (k_x^2+k_y^2) +D_2 k_z^2$. 
Similar to before, the low-energy Hamiltonian in Eq.~\eqref{BiSe_Hamiltonian} can be promoted to a tight-binding Hamiltonian via the minimal lattice regularization;
$\epsilon_\mathbf{k} = C + 2 \tilde{D}_1 [2 - \cos{(k_x a)} - \cos{(k_y a)}] + 2 \tilde{D}_2 [1 - \cos{(k_z a)}]$, 
$d_{\mathbf{k},\pm} = \tilde{A}_1 [\sin{(k_x a)} \pm i \sin{(k_y a)}]$, 
$d_{\mathbf{k},3} = M + 2 \tilde{B}_1 [2 - \cos{(k_x a)} - \cos{(k_y a)}] + 2 \tilde{B}_2 [1 - \cos{(k_z a)}]$, 
$d_{\mathbf{k},4} = \tilde{A}_2 \sin{(k_z a)}$, and
$d_{\mathbf{k},5} = 0$,
where we define $\tilde{A}_i = A_i / a$, $\tilde{B}_i = B_i / a^2$, and $\tilde{D}_i = D_i / a^2$ ($i=1,2$).

%%%%%%%%%%%%%%%%%%%%%%%%%%%%%%%%%%%%%%%%%%%%%%%%%%%%%%%%%%%%%%%%%%%%%%
%%%%%%%%%%%%%%%%%%%%%%%%%%%%%%%%%%%%%%%%%%%%%%%%%%%%%%%%%%%%%%%%%%%%%%
\begin{figure}[t]
\centering
\includegraphics[width=0.45\textwidth]
{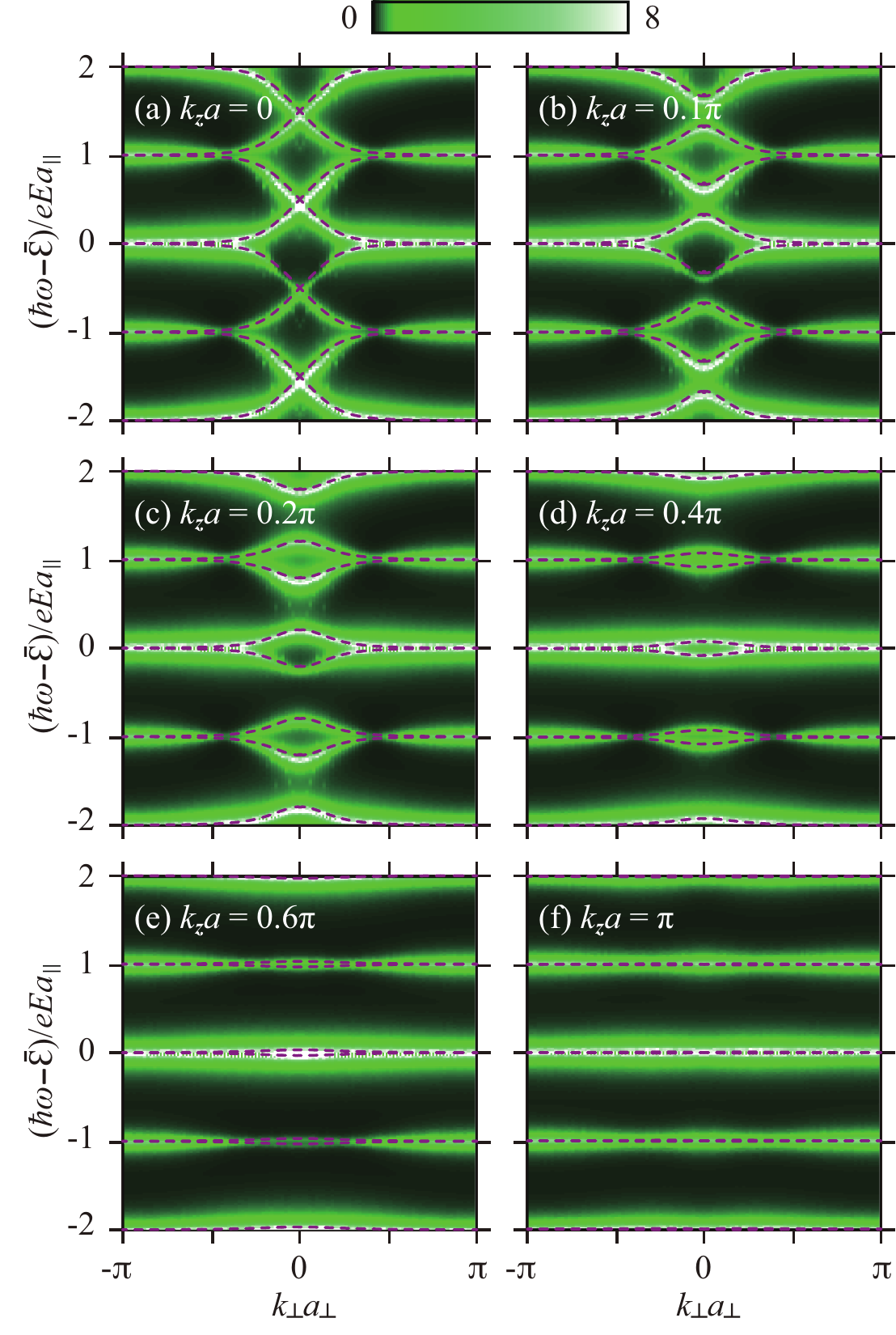} \\
\caption{
Semi-local DOS as a function of $k_\perp$ in the 3D TI model within various 2D subspaces lying parallel to the $k_x\mbox{--}k_y$ plane at different $k_z$. %showing that the 3D topological invariants are directly manifested in the winding numbers of the WSL within the inversion-symmetric 2D subspaces satisfying the Abelian condition.
Here, the semi-local DOS is summed over for all generalized orbitals including both spin and orbital degrees of freedom.
The electric field with magnitude $eEa_\parallel/4\tilde{D}_1 = 0.16$ is applied along the principal lattice direction within the $x\mbox{--}y$ plane, in which situation $a_\parallel=a_\perp=a$.
Model parameters are chosen such that $\tilde{A}_1/4\tilde{D}_1 = 0.6$, $\tilde{A}_2/4\tilde{D}_1 = 0.5$, $\tilde{B}_1/4\tilde{D}_1 = 0.6$, $\tilde{B}_2/4\tilde{D}_1 = 0.3$, $C/4\tilde{D}_1=0$, $\tilde{D}_2/4\tilde{D}_1 = 0.2$, and $M /4\tilde{D}_1 = - 0.3$.
Panels (a) and (f) describe the inversion-symmetric 2D subspaces, where the Berry connection/curvature becomes Abelian and thus the winding number of the WSL is well defined. 
The Kramers doublets exchange partners in panel (a) while not in panel (f), which means that the strong $\mathbb{Z}_2$ invariant is non-trivial.
In general 2D subspaces [panels (b)\mbox{--}(e)], the Berry connections/curvatures are non-Abelian.
The energy spectra of the WSL eigenstates are accurately captured by the dashed guide lines obtained from the requantized effective theory via solving Eq.~\eqref{Heff_eigenvalue_eq}, which covers the Abelian situations in panels (a) and (f) as limiting cases. 
}
\label{Fig6_3DTI_xy}
\end{figure}
%%%%%%%%%%%%%%%%%%%%%%%%%%%%%%%%%%%%%%%%%%%%%%%%%%%%%%%%%%%%%%%%%%%%%%
%%%%%%%%%%%%%%%%%%%%%%%%%%%%%%%%%%%%%%%%%%%%%%%%%%%%%%%%%%%%%%%%%%%%%%

%%%%%%%%%%%%%%%%%%%%%%%%%%%%%%%%%%%%%%%%%%%%%%%%%%%%%%%%%%%%%%%%%%%%%%
%%%%%%%%%%%%%%%%%%%%%%%%%%%%%%%%%%%%%%%%%%%%%%%%%%%%%%%%%%%%%%%%%%%%%%
\begin{figure}[t]
\centering
\includegraphics[width=0.45\textwidth]
{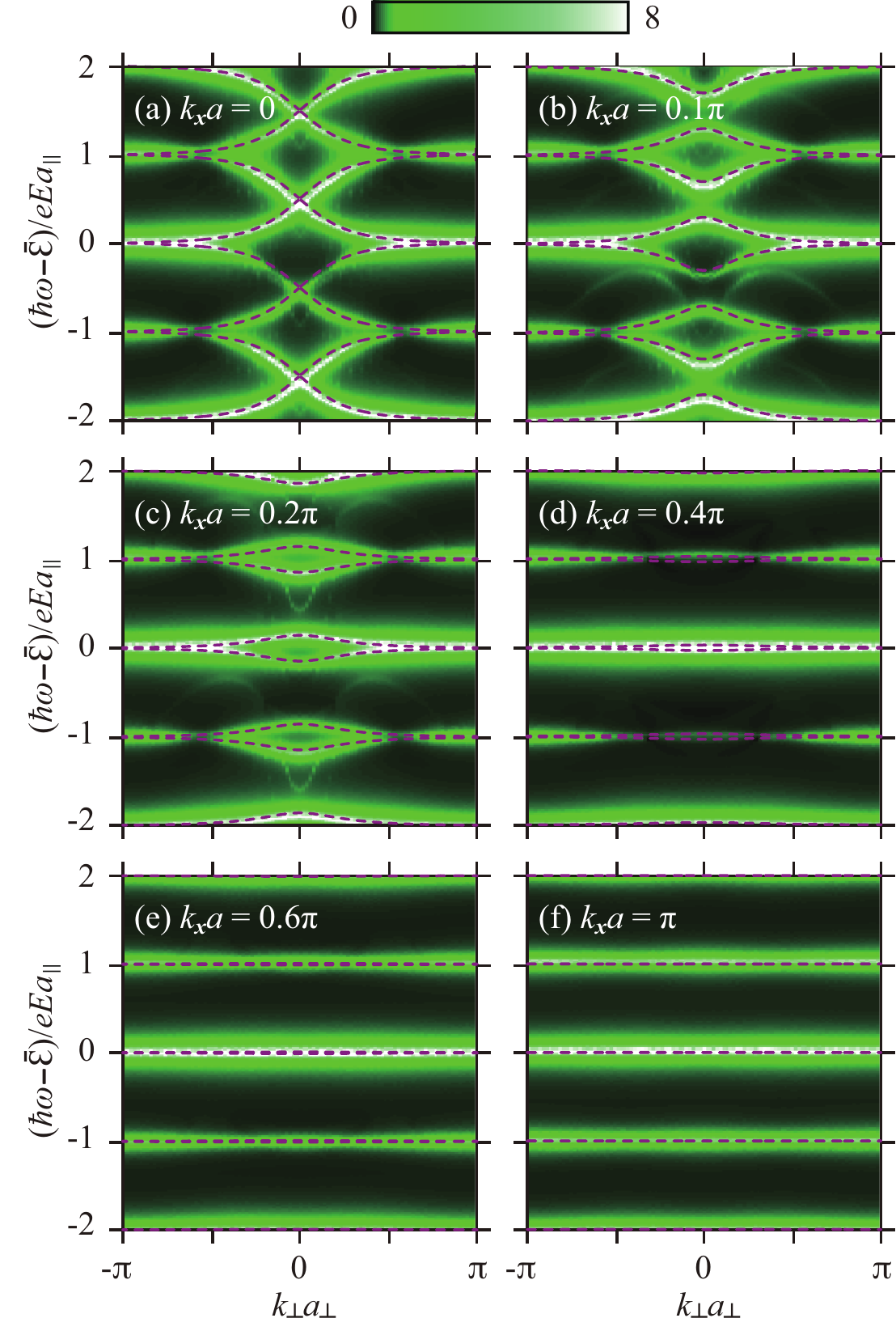} \\
\caption{
Counterpart of Fig.~\ref{Fig6_3DTI_xy} within various 2D subspaces lying parallel to the $k_y\mbox{--}k_z$ plane as a function of $k_x$. 
Model parameters are the same as in Fig.~\ref{Fig6_3DTI_xy}.
}
\label{Fig7_3DTI_yz}
\end{figure}
%%%%%%%%%%%%%%%%%%%%%%%%%%%%%%%%%%%%%%%%%%%%%%%%%%%%%%%%%%%%%%%%%%%%%%
%%%%%%%%%%%%%%%%%%%%%%%%%%%%%%%%%%%%%%%%%%%%%%%%%%%%%%%%%%%%%%%%%%%%%%

As mentioned previously, the 3D band topology is characterized by four different $\mathbb{Z}_2$ invariants, $(\nu_0;\nu_1,\nu_2,\nu_3)$. 
Governing the robustness of a given 3D TI, the strong $\mathbb{Z}_2$ invariant $\nu_0$ becomes non-zero if the 2D topological invariant of the inversion-symmetric 2D subspace containing one set of four TRIM is different from that containing the other set.  
What this means in the cubic lattice is that the 2D topological invariant, or equivalently the winding number of the WSL, in the 2D subspace lying parallel to the $k_x\mbox{--}k_y$ plane at $k_z a=0$ should be opposite to that at $k_z a=\pm\pi$. 
Of course, this statement should be true regardless of whether we choose the $k_x a=0$ and $\pi$ subspaces (or the $k_y a=0$ and $\pi$ subspaces) instead of the $k_z a=0$ and $\pi$ counterparts.

The choice of the $k_z a=0$ and $\pi$ subspaces is particularly convenient since, with $d_{{\bf k},4}=0$ at $k_z a=0$ and $\pm\pi$, the Hamiltonian in Eq.~\eqref{BiSe_Hamiltonian} becomes explicitly block-diagonalized with each $2\times 2$ block precisely reducing to the 2D TI Hamiltonian of the BHZ model in Eq.~\eqref{BHZ_Hamiltonian}.
In this situation, exactly the same Zak phase analysis used for the 2D TI model can be applied to determine the winding number of the WSL in the 2D subspaces at $k_z a=0$ and $\pm\pi$.
Consequently, in the current model, the 3D band topology becomes non-trivial if the winding number of the WSL at $k_z a=0$ is non-zero while that at $k_z a=\pi$ is zero, or vice versa. 
As one can see in Figs.~\ref{Fig6_3DTI_xy}~(a) and (f), this condition can be satisfied for an appropriate set of model parameters, generating a strong 3D TI phase.

Meanwhile, in general 2D subspaces at $k_z a \neq 0, \pm\pi$, the Berry connections/curvatures become non-Abelian. 
In this situation, the Zak phase cannot be properly defined and thus the energy spectrum of the WSL eigenstates is no longer described by the simple semiclassical theory of the Abelian Berry connection/curvature in Sec.~\ref{sec:Semiclassical_theory}. 
Instead, the energy spectrum of the WSL eigenstates is computed via the requantized effective theory of the general non-Abelian Berry connection/curvature in Sec.~\ref{sec:Requantized_effective_theory}.
Figures~\ref{Fig6_3DTI_xy}~(b)\mbox{--}(e) show the comparison between the semi-local DOS obtained from the full quantum theory and the guide lines of the WSL eigenstates (dashed lines) obtained from the requantized effective theory in Sec.~\ref{sec:Requantized_effective_theory}. 
As one can see, the agreement between the two theories is excellent.
It is important to note that the requantized effective theory reproduces the results of the semiclassical theory of the Abelian Barry connection/curvature in Fig.~\ref{Fig6_3DTI_xy}~(a) and (f) in the Abelian limit.

Finally, it is important to check if the strong $\mathbb{Z}_2$ invariant is uniquely determined, being independent of the choice of 2D subspaces.
To this end, we examine the evolution of the semi-local DOS in various 2D subspaces lying parallel to the $k_y\mbox{--}k_z$ plane as a function of $k_x$, which is shown in Fig.~\ref{Fig7_3DTI_yz}.
Unlike in Figs.~\ref{Fig6_3DTI_xy}~(a) and (f), here, the Hamiltonian is not explicitly block-diagonalized even within the inversion-symmetric 2D subspaces at $k_x a=0$ and $\pi$.
Fortunately, as explained in Sec.~\ref{sec:Requantized_effective_theory}, the Berry connections satisfy the Abelian condition in Eq.~\eqref{AbelianCondition} within these 2D subspaces.
This means that, within these 2D subspaces, the Hamiltonian is decomposed into two independent parts with each having the well-defined winding number of the WSL. 
Moreover, the time-reversal symmetry dictates that the two winding numbers should be opposite.
As one can see, this is exactly confirmed in Figs.~\ref{Fig7_3DTI_yz}~(a) and (f).
As before, in general 2D subspaces where the Berry connections/curvatures are non-Abelian, the energy spectrum of the WSL eigenstates is accurately captured by the requantized effective theory in Sec.~\ref{sec:Requantized_effective_theory}. 
A similar analysis for the 2D subspaces lying parallel to the $k_z\mbox{--}k_x$ plane is guaranteed to generate exactly the same conclusion as the above due to the reflection symmetry between the $x$ and $y$ directions.

%%%%%%%%%%%%
%%% Discussion %%%
%%%%%%%%%%%%
\section{Discussion}
\label{sec:Discussion}

In this work, we show that the non-trivial band topologies of both 2D and 3D TIs, characterized by the Chern numbers and the $\mathbb{Z}_2$ invariants, respectively, are directly manifested in the winding numbers of the WSL emerging under an electric field.
Being alternative to the topological magneto-electric effect~\cite{Qi08b}, this provides a spectroscopic method to measure the topological invariants directly in the bulk of both 2D and 3D TIs.
Below, we discuss briefly how this method can be realized in actual experiments.

The main physical observable to be measured is the semi-local DOS. 
Considering that the modern STM technique has the sufficient spatial resolution to distinguish individual atoms in a crystal, 
the semi-local DOS can be in principle obtained by scanning the surface of a 2D TI (possibly, obtained in the thin-film limit of 3D TIs~\cite{Lu10}) or the cleaved surfaces of a 3D TI and then partially Fourier-transforming the STM data along the perpendicular direction to the electric field.
Another method to measure the semi-local DOS is the ARPES, which can give rise to the momentum-resolved information directly near the cleaved surfaces of a 3D TI.

The WSL has been so far observed only in man-made structures such as optical and semiconductor superlattices since the typical lattice constant in a natural crystal is usually too small ($\sim$ a few $\mathring{\rm A}$) that the energy spectrum of the WSL eigenstates (broadened by impurity scattering) is not well resolved for a typical strength of electric field~\cite{Mendez93, Gluck02}. 
A key task is to apply a sufficiently strong electric field to overcome the broadening effect, while suppressing the Joule heating. 
This has been achieved in semiconductor superlattices with the superperiod of 50\mbox{--}100 $\mathring{\rm A}$ at an electric field of 10\mbox{--}20 kV/cm, in which situation the energy spacing between the WSL eigenstates is roughly 10\mbox{--}40 meV~\cite{Mendez93, Gluck02}. 
If experiments can be performed in a natural crystal with the same energy resolution, the required electric field is estimated to be roughly in the order of 100\mbox{--}200 kV/cm.
With stronger electric fields, it would be important to make TIs truly bulk-insulating~\cite{Ren10, Brahlek14,Xiong12} in order to suppress the Joule heating.

Actually, it has been proposed that an artificial TI (as well as an artificial Weyl semimetal) can be constructed in a superlattice structure, which is composed of alternating layers of topological and ordinary insulators with the layer thickness spanning many ($\sim 20\mbox{--}30$) unit cells~\cite{Burkov11}. 
In this situation, the energy spacing between the WSL eigenstates can be dramatically enlarged, opening up the possibility of observing the winding number of the WSL even in a weak electric field.   
For future work, we would like to investigate if this possibility can be realized in realistic material conditions~\cite{Unpublished_Kim}.

Furthermore, it would be also interesting to observe the energy spectrum of the WSL eigenstates in graphene, which is topologically critical as explained in Fig.~\ref{Fig5_KM}.
In graphene, the limitation of a small unit cell can be effectively overcome by forming the moir\'{e} structure~\cite{Yankowitz12, Ponomarenko13, Dean13}. Moreover, extensive efforts are being devoted to enhance the spin-orbit coupling strength~\cite{CastroNeto09, Weeks11, Marchenko12, Balakrishnan13} to realize the KM model in actual graphene.

%%%%%%%%%%%%%%%%
%%% Acknowledgements %%%
%%%%%%%%%%%%%%%%
\acknowledgments
%%%%%%%%%%%%%%%%
The authors are indebted to Seongshik (Sean) Oh, Changyoung Kim, Jhinhwan Lee, Kee Hoon Kim, and Tae Won Noh for their insightful comments on the experimental realization of our work.
Also, the authors are grateful to Kun Woo Kim, Sangmo Cheon, Hyun Woong Kwon, Jun-Won Rhim, Jae-Seung Jeong, and Suk Bum Chung for sharing illuminating discussions. 
The authors thank KIAS Center for Advanced Computation (CAC) for providing computing resources.

%%%%%%%%%%%
%%% Appendix %%%
%%%%%%%%%%%
\appendix
%%%%%%%%%%%

%%%%%%%%%%%%%%%%%%%%%%%%%%%%%%%%%%%%%%
\section{Zak phase}
\label{Appen:Zak}
%%%%%%%%%%%%%%%%%%%%%%%%%%%%%%%%%%%%%%

In this section, we provide computational details of the Zak phase in 2D TIs, where the Berry connection/curvature is Abelian.
The 2D TI Hamiltonian in Eq.~\eqref{Hamiltonian2DTI} is repeated here for convenience:
$H = \sum_{\mathbf{k}, \sigma=\uparrow,\downarrow} \psi_{\mathbf{k}\sigma}^\dag H_\sigma(\mathbf{k}) \psi_{\mathbf{k}\sigma}$ with 
$H_\downarrow(\mathbf{k}) = H_\uparrow^*(-\mathbf{k})$ and 
\begin{align}
H_\uparrow(\mathbf{k})
= \left(
\begin{array}{cc}
\epsilon_\mathbf{k} + d_{\mathbf{k},z}
& d_{\mathbf{k},-} \\
d_{\mathbf{k},+}
& \epsilon_\mathbf{k} - d_{\mathbf{k},z}
\end{array}
\right),
\label{Generic_2DTI_Hamiltonian}
\end{align}
where the detailed form of ${\bf d}_{\bf k}$ is not important for the current purpose except that $|{\bf d}_{\bf k}|=|{\bf d}_{-{\bf k}}|$ due to the time-reversal symmetry.

After diagonalizing Eq.~\eqref{Generic_2DTI_Hamiltonian}, the energy eigenvalue is obtained as follows: 
\begin{align}
\mathcal{E}_\pm(\mathbf{k})
& = \epsilon_\mathbf{k} \pm |\mathbf{d}_\mathbf{k}|, 
\label{EnergyDispersion}
\end{align}
for both spins, indicating that the system becomes insulating when $|\mathbf{d}_\mathbf{k}| \neq 0$ in the entire BZ. 
The corresponding eigenstates for the lower and upper bands, $\phi_{-,\sigma}({\bf k})$ and $\phi_{+,\sigma}({\bf k})$, respectively, are given as follows:
\begin{align}
|\phi_{\pm,\uparrow}(\mathbf{k}) \rangle
& = \frac{1}{\sqrt{1 + (\zeta_{\mathbf{k},\pm})^2}}
\left(
\begin{array}{c}
\zeta_{\mathbf{k},\pm} e^{-i\varphi_\mathbf{k}} \\
1
\end{array}
\right), 
\label{EigenstateSpinUp}
\\
|\phi_{\pm,\downarrow}(\mathbf{k}) \rangle
& = \frac{1}{\sqrt{1 + (\zeta_{\mathbf{k},\pm})^2}}
\left(
\begin{array}{c}
-\zeta_{\mathbf{k},\pm} e^{i\varphi_\mathbf{k}}  \\
1
\end{array}
\right), 
\label{EigenstateSpinDown}
\end{align}
where
$\zeta_{\mathbf{k},\pm} = (d_{\mathbf{k},z} \pm |\mathbf{d}_\mathbf{k}|) / \sqrt{(d_{\mathbf{k},x})^2 + (d_{\mathbf{k},y})^2}$ and $\varphi_\mathbf{k} = \tan^{-1}(d_{\mathbf{k},y} / d_{\mathbf{k},x})$.

Now, with help of Eqs.~\eqref{EigenstateSpinUp} and \eqref{EigenstateSpinDown}, the Berry connections for spin up and down can be computed as follows:
\begin{align}
\mathcal{A}_{\pm,\uparrow}({\bf k})
& = \langle \phi_{\pm,\uparrow}(\mathbf{k})| i \nabla_\mathbf{k} |\phi_{\pm,\uparrow}(\mathbf{k})\rangle \nonumber\\
& = - \alpha_{{\bf k},\pm} (d_{\mathbf{k},y} \nabla_{\bf k} d_{\mathbf{k},x} - d_{\mathbf{k},x} \nabla_{\bf k} d_{\mathbf{k},y}), 
\label{BerryConnectionSpinUp}
\\
\mathcal{A}_{\pm,\downarrow}({\bf k})
& = \langle \phi_{\pm,\downarrow}(\mathbf{k})| i \nabla_\mathbf{k} |\phi_{\pm,\downarrow}(\mathbf{k})\rangle 
= -\mathcal{A}_{\pm,\uparrow}({\bf k})
%\nonumber \\ & = \alpha_{\bf k} (d_{\mathbf{k},y} \nabla_{\bf k} d_{\mathbf{k},x} - d_{\mathbf{k},x} \nabla_{\bf k} d_{\mathbf{k},y}),
\label{BerryConnectionSpinDown}
\end{align}
where
\begin{align}
\alpha_{{\bf k},\pm} = \frac{|\mathbf{d}_\mathbf{k}| \pm d_{\mathbf{k},z}}{2|\mathbf{d}_\mathbf{k}| [(d_{\mathbf{k},x})^2 + (d_{\mathbf{k},y})^2]}.
\end{align}

As discussed in the main text, the energy spectrum of the WSL eigenstates for each spin component $\sigma$ is given by
\begin{align}
{\cal E}^{\rm WSL}_{n, \pm, \sigma}(k_\perp) = \bar{{\cal E}}_{\pm}(k_\perp) +\left( n +\frac{\gamma_{{\rm Zak}, \pm, \sigma}(k_\perp)}{2\pi}  \right) e E a_\parallel, 
\end{align}
where $\bar{{\cal E}}_{\pm}(k_\perp)=\frac{a_\parallel}{2\pi} \oint_C d k_\parallel {\cal E}_{\pm}({\bf k})$.
The spin-dependent Zak phase, $\gamma_{\mathrm{Zak},\pm,\sigma}(k_\perp)$, is evaluated via
\begin{align}
\gamma_{\mathrm{Zak},\pm,\sigma}(k_\perp) 
& = \oint_C d\mathbf{k}_\parallel \cdot \mathcal{A}_{\pm,\sigma}({\bf k}),
\end{align}
which indicates that $\gamma_{\mathrm{Zak},\pm,\uparrow}(k_\perp)=-\gamma_{\mathrm{Zak},\pm,\downarrow}(k_\perp)$ due to Eq.~\eqref{BerryConnectionSpinDown}, which in turn means that the winding numbers are opposite for different spin components. 
%in the topologically non-trivial phase.

It is important to note that the same formalism can be applied to the inversion-symmetric 2D subspaces within the 3D BZ of 3D TIs, where the Berry connections/curvatures are Abelian. 
By choosing the right basis of wave functions, the 3D TI Hamiltonian can be written as Eq.~\eqref{Generic_2DTI_Hamiltonian} with conserved {\it pseudospin} components.

%%%%%%%%%%%%%%%%%%%
\section{Proof of the Abelian condition} 
\label{Appen:Abelian_condition}
%%%%%%%%%%%%%%%%%%%

In this appendix, we provide the proof of the Abelian condition in Eq.~\eqref{AbelianCondition} for the generic 3D TI model.
To this end, let us rewrite the generic 3D TI Hamiltonian in Eq.~\eqref{Hamiltonian3DTI}:
$H = \sum_{\mathbf{k}} \psi_\mathbf{k}^\dag H(\mathbf{k}) \psi_\mathbf{k}$ with
\begin{align}
H(\mathbf{k})
= \left(
\begin{array}{cccc}
\epsilon_\mathbf{k} - d_{\mathbf{k},3} & d_{\mathbf{k},4} & 0 & d_{\mathbf{k},-} \\
d_{\mathbf{k},4} & \epsilon_\mathbf{k} + d_{\mathbf{k},3} & d_{\mathbf{k},-} & 0 \\
0 & d_{\mathbf{k},+} & \epsilon_\mathbf{k} - d_{\mathbf{k},3} & - d_{\mathbf{k},4} \\
d_{\mathbf{k},+} & 0 & - d_{\mathbf{k},4} & \epsilon_\mathbf{k} + d_{\mathbf{k},3}
\end{array}
\right),
\label{Generic_3DTI_Hamiltonian}
\end{align}
where, via the minimal lattice regularization of the low-energy effective model, the system parameters can be written as 
$\epsilon_\mathbf{k} = C + 2 \tilde{D}_1 [2 - \cos{(k_x a)} - \cos{(k_y a)}] + 2 \tilde{D}_2 [1 - \cos{(k_z a)}]$, 
$d_{\mathbf{k},\pm} = \tilde{A}_1 [\sin{(k_x a)} \pm i \sin{(k_y a)}]$, 
$d_{\mathbf{k},3} = M + 2 \tilde{B}_1 [2 - \cos{(k_x a)} - \cos{(k_y a)}] + 2 \tilde{B}_2 [1 - \cos{(k_z a)}]$, 
$d_{\mathbf{k},4} = \tilde{A}_2 \sin{(k_z a)}$, and
$d_{\mathbf{k},5} = 0$.

Similar to the 2D TI case, the energy eigenvalue is given by
\begin{align}
\mathcal{E}_\pm(\mathbf{k})
= \epsilon_\mathbf{k} \pm |\mathbf{d}_\mathbf{k}| ,
\end{align}
which indicates that there is a double degeneracy for both upper and lower bands.
Let us distinguish the two degenerate energy eigenstates within the upper (subscript $+$) and lower (subscript $-$) bands by introducing a pseudospin index, say, $u$ and $d$.
In other words, the energy eigenstates are distinguished by two indices with one being $\pm$ and the other being $u/d$.
Specifically, the energy eigenstates, $|\phi_{\pm,u}({\bf k})\rangle$ and $|\phi_{\pm,d}({\bf k})\rangle$, are given as follows:
\begin{align}
|\phi_{\pm,u}(\mathbf{k}) \rangle
& = \frac{1}{\sqrt{1 + (\chi_\mathbf{k})^2 + (\zeta_{\mathbf{k}, \pm})^2}}
\left(
\begin{array}{c}
\chi_\mathbf{k} e^{-i\varphi_\mathbf{k}} \\
\zeta_{\mathbf{k},\pm} e^{-i\varphi_\mathbf{k}} \\
1 \\
0
\end{array}
\right),
\label{Eigen3Da}
\\
|\phi_{\pm,d}(\mathbf{k}) \rangle
& = \frac{1}{\sqrt{1 + (\chi_\mathbf{k})^2 + (\zeta_{\mathbf{k},\pm})^2}}
\left(
\begin{array}{c}
1 \\
0 \\
- \chi_\mathbf{k} e^{i\varphi_\mathbf{k}} \\
\zeta_{\mathbf{k},\pm} e^{i\varphi_\mathbf{k}}
\end{array}
\right),
\label{Eigen3Db}
\end{align}
where
$\chi_\mathbf{k} = d_{\mathbf{k},4} / \sqrt{(d_{\mathbf{k},1})^2 + (d_{\mathbf{k},2})^2}$, 
$\zeta_{\mathbf{k},\pm} = (d_{\mathbf{k},3} \pm |\mathbf{d}_\mathbf{k}|) / \sqrt{(d_{\mathbf{k},1})^2 + (d_{\mathbf{k},2})^2}$, and 
$\varphi_\mathbf{k} = \tan^{-1}(d_{\mathbf{k},2}/d_{\mathbf{k},1})$.

In general, the Berry connection has off-diagonal matrix elements mixing between $|\phi_{\pm,u}({\bf k})\rangle$ and $|\phi_{\pm,d}({\bf k})\rangle$, which generates the SU(2) non-Abelian gauge structure~\cite{Wilczek84, Shindou05, Culcer05, Chang08, Xiao10}.
Specifically, the Berry connection for the lower band is explicitly written as follows:
\begin{align}
\mathcal{A}_{\mathbf{k},uu}
= &\langle \phi_{-,u}(\mathbf{k})| i \nabla_\mathbf{k} |\phi_{-,u}(\mathbf{k})\rangle \nonumber\\
= &- \alpha_{\bf k} (d_{\mathbf{k},2} \nabla_{\bf k} d_{\mathbf{k},1} - d_{\mathbf{k},1} \nabla_{\bf k} d_{\mathbf{k},2}),
\label{A_uu}
\\
\mathcal{A}_{\mathbf{k},dd}
= &\langle \phi_{-,d}(\mathbf{k})| i \nabla_\mathbf{k} |\phi_{-,d}(\mathbf{k})\rangle \nonumber\\
= &\alpha_{\bf k} (d_{\mathbf{k},2} \nabla_{\bf k} d_{\mathbf{k},1} - d_{\mathbf{k},1} \nabla_{\bf k} d_{\mathbf{k},2}),
\label{A_dd}
\\
\mathcal{A}_{\mathbf{k},ud} 
= &\mathcal{A}^*_{\mathbf{k},du} 
= \langle \phi_{-,u}(\mathbf{k})| i \nabla_\mathbf{k} |\phi_{-,d}(\mathbf{k})\rangle \nonumber\\
= &\beta_{\bf k} \big[ d_{\mathbf{k},4} ( - i \nabla_{\bf k} d_{\mathbf{k},1} - \nabla_{\bf k} d_{\mathbf{k},2}) \nonumber\\
&+ (i d_{\mathbf{k},1} + d_{\mathbf{k},2}) \nabla_{\bf k} d_{\mathbf{k},4} \big],
\label{A_ud}
\end{align}
where
\begin{align}
\alpha_{\bf k} 
& = \frac{(|\mathbf{d}_\mathbf{k}| - d_{\mathbf{k},3})^2 + (d_{{\bf k},4})^2}{2|\mathbf{d}_\mathbf{k}| (|\mathbf{d}_\mathbf{k}| - d_{\mathbf{k},3}) [(d_{\mathbf{k},1})^2 + (d_{\mathbf{k},2})^2]},
\\
\beta_{\bf k} 
& = \frac{- (d_{\mathbf{k},1} + i d_{\mathbf{k},2})^2}{2|\mathbf{d}_\mathbf{k}| (|\mathbf{d}_\mathbf{k}| - d_{\mathbf{k},3}) [(d_{\mathbf{k},1})^2 + (d_{\mathbf{k},2})^2]}.
\end{align}

Now, one can rearrange Eqs.~\eqref{A_uu}, \eqref{A_dd}, and \eqref{A_ud} as follows:
\begin{align}
\mathcal{A}_{\bf k}^x 
= & \beta_{\bf k} \tilde{A}_1 \tilde{A}_2 a \sin(k_z a) \cos(k_x a) \sigma_y 
\nonumber \\
& - \alpha_{\bf k} \tilde{A}_1^2 a \sin(k_y a) \cos(k_x a) \sigma_z,
\\
\mathcal{A}_{\bf k}^y 
= & - \beta_{\bf k} \tilde{A}_1 \tilde{A}_2 a \sin(k_z a) \cos(k_y a) \sigma_x 
\nonumber \\
& + \alpha_{\bf k} \tilde{A}_1^2 a \sin(k_x a) \cos(k_y a) \sigma_z,
\\
\mathcal{A}_{\bf k}^z 
= & \beta_{\bf k} \tilde{A}_1 \tilde{A}_2 a \sin(k_y a) \cos(k_z a) \sigma_x 
\nonumber \\
& - \beta_{\bf k} \tilde{A}_1 \tilde{A}_2 a \sin(k_x a) \cos(k_z a) \sigma_y,
\end{align}
where the Pauli matrices, $(\sigma_x, \sigma_y, \sigma_z)$, are represented in the basis of $|\phi_{-,u}({\bf k}) \rangle$ and $|\phi_{-,d}({\bf k}) \rangle$ as follows:
\begin{align}
\sigma_x &= |\phi_{-,u}({\bf k}) \rangle \langle \phi_{-,d}({\bf k})| + |\phi_{-,d}({\bf k}) \rangle \langle \phi_{-,u}({\bf k})| , 
\\
\sigma_y &= -i |\phi_{-,u}({\bf k}) \rangle \langle \phi_{-,d}({\bf k})| + i |\phi_{-,d}({\bf k}) \rangle \langle \phi_{-,u}({\bf k})| ,  
\\
\sigma_z &= |\phi_{-,u}({\bf k}) \rangle \langle \phi_{-,u}({\bf k})| - |\phi_{-,d}({\bf k}) \rangle \langle \phi_{-,d}({\bf k}) |.
\end{align}

Then, after some algebra, one can show that the commutator between various components of the Berry connection is summarized compactly as follows: 
\begin{align}
[\mathcal{A}_{\bf k}^i, \mathcal{A}_{\bf k}^j] = 2i\epsilon_{ijk} \sin{(k_k a)} \cos{(k_i a)} \cos{(k_j a)} {\cal F}_{\bf k} ,
\label{BerryConnectionCommutator}
\end{align}
where 
\begin{align}
{\cal F}_{\bf k} = & \alpha_{\bf k} \beta_{\bf k} \tilde{A}_1^3 \tilde{A}_2 a^2   \left[ \sin{(k_x a)} \sigma_x + \sin{(k_y a)} \sigma_y \right]
\nonumber \\
& + \beta_{\bf k}^2 \tilde{A}_1^2 \tilde{A}_2^2 a^2 \sin{(k_z a)} \sigma_z .
\end{align}
Equation~\eqref{BerryConnectionCommutator} implies that the commutator vanishes in the entire $k_i\mbox{--}k_j$ plane if $k_k = 0, \pm \pi/a$, which is nothing but the Abelian condition for the inversion-symmetric 2D subspaces. 
This completes the proof of the Abelian condition in Eq.~\eqref{AbelianCondition}.

%%%%%%%%%%%%%%%%%%%%%%%%%%%%%%%
\section{Floquet Green's function}
\label{Appen:FloquetGreensFunc}
%%%%%%%%%%%%%%%%%%%%%%%%%%%%%%%

The goal of this appendix is to explain how to compute the Floquet Green's function, which satisfies the following equation:
\begin{align}
&\sum_{l,\gamma} 
\left\{ \left[\hbar(\omega + n\tilde{\Omega}) + i\eta \right] \delta_{\alpha\gamma}\delta_{nl}  - H^{nl}_{\alpha\gamma}(\mathbf{k}) \right\} 
(G^r_{\bf k})_{\gamma\beta}^{lm}(\omega)
\nonumber \\
&\;\;\;\;\;= \delta_{\alpha\beta} \delta_{nm},
\label{Appen:FloquetGreensFuncEq}
\end{align}
where $H^{nm}_{\alpha\beta}({\bf k}) = \frac{1}{T} \int_0^T dt e^{i(n-m)\tilde{\Omega} t} H_{\alpha\beta}( \mathbf{k} - e\mathbf{E}t/\hbar )$ with $T = 2\pi/\tilde{\Omega}$.
Since the concrete form of the Hamiltonian matrix element $H_{\alpha\beta}({\bf k})$ depends on the specific model, it is not possible to obtain the general solution for the Floquet Green's function in a closed analytic form.
Fortunately, considering the structure of the generic TI Hamiltonian, it is possible to derive a formal solution for the Floquet Green's function with the same orbital indices, $(G^r_{\bf k})^{nm}_{\alpha\alpha}$, by summing away all other contributions from those with different orbital indices. 
This formal solution is convenient since the semi-local DOS is solely dependent on the orbital-diagonal components $(G^r_{\bf k})^{nn}_{\alpha\alpha}$.
Below, we present such a formal solution first in the case of the 2D TI with conserved spin components and then in the case of 3D TI with mixed spin components.

\subsection{2D TI with conserved spin components}

We begin by rewriting the 2D TI Hamiltonian in a matrix form including both spin components:
\begin{align}
H(\mathbf{k})
= \left(
\begin{array}{cccc}
\epsilon_\mathbf{k} + d_{\mathbf{k},z} & d_{\mathbf{k},-} & 0 & 0 \\
d_{\mathbf{k},+} & \epsilon_\mathbf{k} - d_{\mathbf{k},z} & 0 & 0 \\
0 & 0 & \epsilon_{-\mathbf{k}} + d_{-\mathbf{k},z} & - d_{-\mathbf{k},+} \\
0 & 0 & - d_{-\mathbf{k},-} & \epsilon_{-\mathbf{k}} - d_{-\mathbf{k},z}
\end{array}
\right),
\label{AppenEq:2DTI_Hamiltonian}
\end{align}
where the concrete forms of $\epsilon_{\bf k}$ and ${\bf d}_{\bf k}=(d_{{\bf k},x}, d_{{\bf k},y}, d_{{\bf k},z})$ depend on the specific model and are shown in Sec.~\ref{sec:2D_TI}.
As before, $d_{{\bf k}, \pm}=d_{{\bf k},x} \pm i d_{{\bf k},y}$.
Here, the basis is chosen such that $(1,2,3,4)=(E_1\!\uparrow, H_1\!\uparrow, E_1\!\downarrow, H_1\!\downarrow)$ for the BHZ model and $(A\!\uparrow, B\!\uparrow, A\!\downarrow, B\!\downarrow)$ for the KM model.

By summing away all contributions from the orbital-off-diagonal components in Eq.~\eqref{Appen:FloquetGreensFuncEq}, the inverse of the orbital-diagonal components of the Floquet Green's function can be written in a compact notation with $[{\bf G}_{{\bf k},\alpha\alpha}]^{nm}=(G_{\bf k})^{nm}_{\alpha\alpha}$ as follows:
\begin{align}
{\bf G}^{r-1}_{\mathbf{k},11}(\omega)
& = [{\bf P}^{+}_{\mathbf{k}}(\omega)]^{-1} - {\bf D}_{\mathbf{k}}^{-} \cdot {\bf P}^{-}_{\mathbf{k}}(\omega) \cdot {\bf D}_{\mathbf{k}}^{+},
\label{2DGreen11}
\\
{\bf G}^{r-1}_{\mathbf{k},22}(\omega)
& = [{\bf P}^{-}_{\mathbf{k}}(\omega)]^{-1} - {\bf D}_{\mathbf{k}}^{+} \cdot {\bf P}^{+}_{\mathbf{k}}(\omega) \cdot {\bf D}_{\mathbf{k}}^{-},
\label{2DGreen22}
\\
{\bf G}^{r-1}_{\mathbf{k},33}(\omega)
& = [{\bf P}^{+}_{-\mathbf{k}}(\omega)]^{-1} - {\bf D}_{-\mathbf{k}}^{+} \cdot {\bf P}^{-}_{-\mathbf{k}}(\omega) \cdot {\bf D}_{-\mathbf{k}}^{-},
\label{2DGreen33}
\\
{\bf G}^{r-1}_{\mathbf{k},44}(\omega)
& = [{\bf P}^{-}_{-\mathbf{k}}(\omega)]^{-1} - {\bf D}_{-\mathbf{k}}^{-} \cdot {\bf P}^{+}_{-\mathbf{k}}(\omega) \cdot {\bf D}_{-\mathbf{k}}^{+},
\label{2DGreen44}
\end{align}
where the Floquet matrices ${\bf P}^{\pm}_{\bf k}(\omega)$ and ${\bf D}^{\pm}_{\bf k}$ are given as
\begin{align}
[{\bf P}^{\pm}_{\mathbf{k}}(\omega)]^{-1, nm} = [\hbar(\omega + n\tilde{\Omega}) &+ i \eta] \delta_{nm} 
- (\epsilon_\mathbf{k}^{nm} \pm d_{\mathbf{k},z}^{nm}),
\label{2DGreenP}
\end{align}
and
\begin{align}
({\bf D}_{\mathbf{k}}^{\pm})^{nm} =& d_{\mathbf{k},\pm}^{nm},
\label{2DGreenD}
\end{align}
where
$\epsilon_{\bf k}^{nm} = \frac{1}{T} \int_0^T dt e^{i(m-n)\tilde{\Omega} t} \epsilon( \mathbf{k} - e\mathbf{E}t/\hbar )$ and 
${\bf d}_{\bf k}^{nm} = \frac{1}{T} \int_0^T dt e^{i(m-n)\tilde{\Omega} t} {\bf d}( \mathbf{k} - e\mathbf{E}t/\hbar )$ with $T = 2\pi/\tilde{\Omega}$.

The specific forms of $\epsilon_{\bf k}^{mn}$ and ${\bf d}_{\bf k}^{mn}$ depend on not only the detailed ${\bf k}$ dependences of $\epsilon_{\bf k}$ and ${\bf d}_{\bf k}$, but also the electric-field direction.
In what follows, we present the specific forms of $\epsilon_{\bf k}^{mn}$ and ${\bf d}_{\bf k}^{mn}$ for the BHZ model along the principal direction of the square lattice and the KM model along the armchair as well as the zigzag directions.

\subsubsection{BHZ model}

The BHZ model is defined on the square lattice.
To appreciate that the concrete form of the Floquet matrices depends on the electric-field direction, let us imagine that the electric field is applied along the direction with angle $\theta$ measured from the principal, say, $x$ axis of the square lattice. In this situation, the Peierls-shifted crystal momentum becomes 
\begin{align}
&\left(
\begin{array}{c}
\hbar k_x + \frac{e}{c} A_x(t) \\
\hbar k_y + \frac{e}{c} A_y(t)
\end{array}
\right)
%\nonumber \\
= \left(
\begin{array}{cc}
\cos\theta & -\sin\theta \\
\sin\theta & \cos\theta
\end{array}
\right) \left(
\begin{array}{c}
\hbar k_\parallel - eEt \\
\hbar k_\perp
\end{array}
\right) \nonumber\\
&= \frac{1}{\sqrt{p^2 + q^2}} \left(
\begin{array}{c}
(q \hbar k_\parallel - p \hbar k_\perp) -qeEt \\
(p \hbar k_\parallel +q \hbar k_\perp) -peEt
\end{array}
\right),
\end{align}
%}
%\end{widetext}
where we set $\theta = \tan^{-1}(p/q)$ with $p,q \in \mathbb{Z}$ so that all different time-dependent terms in $\epsilon({\bf k}-e{\bf E}t/\hbar)$ and ${\bf d}({\bf k}-e{\bf E}t/\hbar)$ become commensurate with each other.
In other words, the entire time dependence occurs through $\cos{(k_x a-eaA_x(t)/\hbar c)}$, $\cos{(k_y a-eaA_y(t)/\hbar c)}$, and their sine counterparts, which means that there exist two different oscillation frequencies: $qeEa/\sqrt{q^2+p^2}$ and $peEa/\sqrt{q^2+p^2}$.
In this situation, the Floquet frequency, which is the natural frequency of the Peierls-shifted Hamiltonian, is given as $\tilde{\Omega} = eE\tilde{a}_\parallel/\hbar$ with $\tilde{a}_\parallel / a = \mathrm{gcd}(q,p) / \sqrt{q^2 + p^2}$, where $\textrm{gcd}(q,p)$ denotes the greatest common divisor of $q$ and $p$. 
Note that, for general angle $\theta$, $\tilde{\Omega}$ is not necessarily the same as the Bloch oscillation frequency $\Omega=eEa_\parallel/\hbar$.

In the case of $(q,p) = (1,0)$, where $\tilde{a}_\parallel=a_\parallel=a$ and thus $\tilde{\Omega}=\Omega$, the Floquet matrices are given as follows:
%\begin{widetext}
\begin{align}
\epsilon_\mathbf{k}^{nm}
&= e^{i (n-m) k_\parallel a} 
\big[ C + 2\tilde{D} (2 - \cos( k_\perp a )) \delta_{nm} 
\nonumber \\
&\;\;\;\;\;- \tilde{D} (\delta_{n,m+1} + \delta_{n,m-1}) \big],
\label{FloquetMatrix1}
\\
d_{\mathbf{k},\pm}^{mn} 
&= \frac{1}{2i} e^{i (n-m) k_\parallel a} \tilde{A}
\big[ \pm 2 \sin( k_\perp a ) \delta_{nm} 
\nonumber \\
&\;\;\;\;\;+ \delta_{n,m+1} - \delta_{n,m-1} \big],
\label{FloquetMatrix2}
\\
d_{\mathbf{k},z}^{mn} 
&= e^{i (n-m) k_\parallel a} 
\big\{ \big[ M + 2\tilde{B} \left( 2 - \cos( k_\perp a ) \right) \big] \delta_{nm} 
\nonumber \\
&\;\;\;\;\;- \tilde{B} (\delta_{n,m+1} + \delta_{n,m-1}) \big\} ,
\label{FloquetMatrix3}
\end{align}
%\end{widetext}
which can be plugged into Eqs.~\eqref{2DGreenP} and \eqref{2DGreenD} to compute the inverses of the Floquet Green's functions in Eqs~\eqref{2DGreen11}\mbox{--}\eqref{2DGreen44}, which are then inverted to generate the Floquet Green's functions themselves.

\subsubsection{KM model}

The KM model is defined on the honeycomb lattice, where the Peierls-shifted crystal momentum is given by
\begin{align}
&\left(
\begin{array}{c}
\hbar k_x + \frac{e}{c} A_x(t) \\
\hbar k_y + \frac{e}{c} A_y(t)
\end{array}
\right) 
= \left(
\begin{array}{cc}
\cos\theta & -\sin\theta \\
\sin\theta & \cos\theta
\end{array}
\right) \left(
\begin{array}{c}
\hbar k_\parallel - eEt \\
\hbar k_\perp
\end{array}
\right) \nonumber\\
&= \frac{1}{\sqrt{q^2 + p^2/3}} \left(
\begin{array}{c}
q (\hbar k_\parallel - eEt) - p k_\perp / \sqrt{3} \\
p (\hbar k_\parallel - eEt) / \sqrt{3} + q k_\perp
\end{array}
\right) ,
\end{align}
%\end{widetext}
where we set $\theta = \tan^{-1}\left(\frac{p/\sqrt{3}}{q}\right)$ with $p,q \in \mathbb{Z}$ so that $(q,p)=(1,0)$ and $(1,1)$ correspond to when the electric field is applied along the armchair and the zigzag directions, respectively. 
Note that $(q,p)=(0,1)$ also corresponds to the zigzag direction. 
The entire time dependence of the Peierls-shifted Hamiltonian occurs through six terms; 
$\exp{(i \mathbf{k}(t) \cdot \mathbf{c}_1)}$, $\exp{(i \mathbf{k}(t) \cdot \mathbf{c}_2)}$, $\exp{(i \mathbf{k}(t) \cdot \mathbf{c}_3)}$,
$\sin{(\mathbf{k}(t) \cdot \mathbf{a}_1)}$,  $\sin{(\mathbf{k}(t) \cdot \mathbf{a}_2)}$, and $\sin{(\mathbf{k}(t) \cdot (\mathbf{a}_1 - \mathbf{a}_2))}$, 
where $\mathbf{c}_1 = a (1/2, \sqrt{3}/2)$, $\mathbf{c}_2 = a (1/2, -\sqrt{3}/2)$, $\mathbf{c}_3 = a (-1, 0)$, $\mathbf{a}_1 = a (3/2, \sqrt{3}/2)$, and $\mathbf{a}_2 = a (3/2, -\sqrt{3}/2)$. 
Note that ${\bf k}(t)={\bf k}-e{\bf E}t/\hbar$.
This means that there are six different oscillation frequencies.
As a consequence, the Floquet frequency is given by $\tilde{\Omega} = eE\tilde{a}_\parallel/\hbar$ with
$\tilde{a}_\parallel / a = \mathrm{gcd}(3q+p, 3q-p, q+p, q-p,  2q, 2p) / 2\sqrt{q^2 + p^2 / 3}$, where ${\rm gcd}(a_1,\cdots, a_n)$ denotes the greatest common divisor among $(a_1, \cdots, a_n)$.

%%%%%%%%%%%%%%%%%%%%%%
\begin{figure}[t]
\centering
\includegraphics[width=0.45\textwidth]
{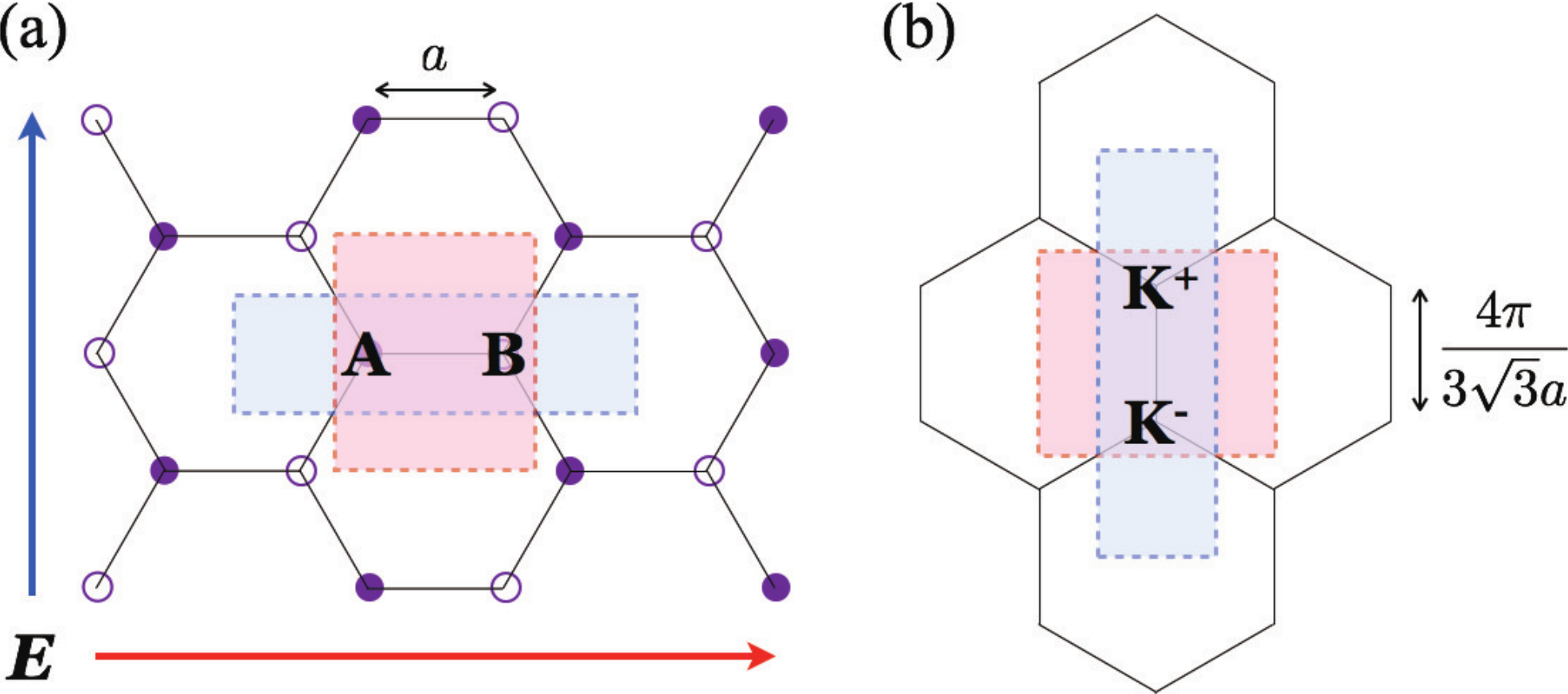} \\
\caption{
Unit cells and the corresponding Brillouin zones in the electric-field-applied KM model.
(a) The red and blue boxes denote the unit cells in the lattice when the electric field is applied along the armchair (red arrow) and the zigzag (blue arrow) direction, respectively.
Here, $A$ and $B$ denote different sublattices.
(b) The corresponding Brillouin zones are shown as the red and blue boxes for the armchair and the zigzag directions, respectively. 
Here, ${\bf K}^\pm=(\frac{2\pi}{3a},\pm\frac{2\pi}{3\sqrt{3}a})$ denote the Dirac points.  
}
\label{Fig8_Appendix}
\end{figure}
%%%%%%%%%%%%%%%%%%%%%%

In the KM model, $\epsilon_{\bf k}^{nm}$ is always zero.
In this work, $d^{nm}_{{\bf k},\pm}$ and $d^{nm}_{{\bf k},z}$ are computed in two different situations with the electric field applied along the armchair and the zigzag directions.
First, the armchair direction is obtained by choosing $(q,p) = (1,0)$, where $\tilde{a}_\parallel=a/2$, while $a_\parallel = 3a/2$ and $a_\perp = \sqrt{3}a$. 
In this situation, the Floquet matrices $d^{nm}_{{\bf k},\pm}$ and $d^{nm}_{{\bf k},z}$ are given by
%\begin{widetext}
\begin{align}
d_{\mathbf{k},\pm}^{nm} 
& = e^{ i (n-m) k_\parallel \tilde{a}_\parallel} \tilde{t} \left[ - 2 \cos{(k_\perp a_\perp / 2)} \delta_{n, m \mp 1} - \delta_{n, m \pm 2} \right],
\\
d_{\mathbf{k},z}^{nm} 
& = e^{ i (n-m) k_\parallel \tilde{a}_\parallel} 
\lambda_{\rm SO}
[ -2 \sin{( k_\perp a_\perp )} 
\delta_{nm} 
\nonumber \\
&\;\;\;\;\;+ 2 \sin{(k_\perp a_\perp / 2)} ( \delta_{n,m+3} + \delta_{n,m-3} ) ].
\end{align}
%\end{widetext}
Second, the zigzag direction is obtained by choosing $(q,p) = (0,1)$, where $\tilde{a}_\parallel=a_\parallel = \sqrt{3}a/2$ and $a_\perp = 3a$. 
In this situation, the Floquet matrices are given by
%\begin{widetext}
\begin{align}
d_{\mathbf{k},\pm}^{nm} 
& = e^{ i (n-m) k_\parallel \tilde{a}_\parallel } \tilde{t} [  - \exp{(\mp i k_\perp a_\perp / 3)} \delta_{nm} 
\nonumber \\
&\;\;\;\;\;- \exp{(\pm i k_\perp a_\perp / 6)} (\delta_{n,m+1} + \delta_{n,m-1}) ],
\\
d_{\mathbf{k},z}^{nm} 
& = e^{ i (n-m) k_\parallel \tilde{a}_\parallel } 
\lambda_{\rm SO}
[ - 2 i \cos{(k_\perp a_\perp / 2)} (\delta_{n,m+1} - \delta_{n,m-1}) 
\nonumber \\
&\;\;\;\;\;+ i (\delta_{n,m+2} - \delta_{n,m-2}) ].
\end{align}
%\end{widetext}
Note that the unit cells for the armchair and the zigzag directions have different shapes, but the same area. 
See Fig.~\ref{Fig8_Appendix} for illustration.

\subsection{3D TI with mixed spin components}

Let us begin by rewriting the 3D TI Hamiltonian in a matrix form:
\begin{align}
H(\mathbf{k})
= \left(
\begin{array}{cccc}
\epsilon_\mathbf{k} - d_{\mathbf{k},3} & d_{\mathbf{k},4} & 0 & d_{\mathbf{k},-} \\
d_{\mathbf{k},4} & \epsilon_\mathbf{k} + d_{\mathbf{k},3} & d_{\mathbf{k},-} & 0 \\
0 & d_{\mathbf{k},+} & \epsilon_\mathbf{k} - d_{\mathbf{k},3} & - d_{\mathbf{k},4} \\
d_{\mathbf{k},+} & 0 & - d_{\mathbf{k},4} & \epsilon_\mathbf{k} + d_{\mathbf{k},3}
\end{array}
\right),
\label{AppenEq:3DTI_Hamiltonian}
\end{align}
where the basis is chosen such that $(1,2,3,4)=(P1_z^{+}\!\uparrow, P2_z^{-}\!\uparrow, P1_z^{+}\!\downarrow, P2_z^{-}\!\downarrow)$.
The concrete forms of $\epsilon_{\bf k}$ and ${\bf d}_{\bf k}$ are shown in Sec.~\ref{sec:3D_TI}.

Similar to the 2D TI case, by summing away all contributions from the orbital-off-diagonal components in Eq.~\eqref{Appen:FloquetGreensFuncEq}, the inverse of the orbital-diagonal components of the Floquet Green's function can be written as follows:
\begin{widetext}
\begin{align}
\mathbf{G}^{r-1}_{\mathbf{k},11}(\omega)
&= [\mathbf{P}^{-}_{\mathbf{k}}(\omega)]^{-1} - \mathbf{D}_{\mathbf{k}}^{-} \cdot \mathbf{Q}_{\mathbf{k}}^{a}(\omega) \cdot \mathbf{D}_{\mathbf{k}}^{+} 
\nonumber \\
&- \Big( \mathbf{D}_{\mathbf{k}}^{4} - \mathbf{D}_{\mathbf{k}}^{-} \cdot \mathbf{Q}_{\mathbf{k}}^{a}(\omega) \cdot \mathbf{D}_{\mathbf{k}}^{4} \cdot \mathbf{P}_{\mathbf{k}}^{-}(\omega) \cdot \mathbf{D}_{\mathbf{k}}^{+} \Big) 
\cdot \mathbf{R}_{\mathbf{k}}^{a+}(\omega) \cdot 
\Big( \mathbf{D}_{\mathbf{k}}^{4} - \mathbf{D}_{\mathbf{k}}^{-} \cdot \mathbf{P}_{\mathbf{k}}^{-}(\omega) \cdot \mathbf{D}_{\mathbf{k}}^{4} \cdot \mathbf{Q}_{\mathbf{k}}^{a}(\omega) \cdot \mathbf{D}_{\mathbf{k}}^{+} \Big),
\label{3DGreen11}
\\
\mathbf{G}^{r-1}_{\mathbf{k},22}(\omega)
&= [\mathbf{P}_{\mathbf{k}}^{+}(\omega)]^{-1} 
- \mathbf{D}_{\mathbf{k}}^{4} \cdot \mathbf{Q}_{\mathbf{k}}^{b+}(\omega) \cdot \mathbf{D}_{\mathbf{k}}^{4} 
\nonumber \\
&- \Big( \mathbf{D}_{\mathbf{k}}^{-} - \mathbf{D}_{\mathbf{k}}^{4} \cdot \mathbf{Q}_{\mathbf{k}}^{b+}(\omega) \cdot \mathbf{D}_{\mathbf{k}}^{-} \cdot \mathbf{P}_{\mathbf{k}}^{+}(\omega) \cdot \mathbf{D}_{\mathbf{k}}^{4} \Big) 
\cdot \mathbf{R}_{\mathbf{k}}^{b+}(\omega) \cdot 
\Big( \mathbf{D}_{\mathbf{k}}^{+} - \mathbf{D}_{\mathbf{k}}^{4} \cdot \mathbf{P}_{\mathbf{k}}^{+}(\omega) \cdot \mathbf{D}_{\mathbf{k}}^{+} \cdot \mathbf{Q}_{\mathbf{k}}^{b+}(\omega) \cdot \mathbf{D}_{\mathbf{k}}^{4} \Big),
\label{3DGreen22}
\\
\mathbf{G}^{r-1}_{\mathbf{k},33}(\omega)
&= [\mathbf{P}^{-}_{\mathbf{k}}(\omega)]^{-1} - \mathbf{D}_{\mathbf{k}}^{+} \cdot \mathbf{Q}_{\mathbf{k}}^{a}(\omega) \cdot \mathbf{D}_{\mathbf{k}}^{-} 
\nonumber \\
&- \Big( \mathbf{D}_{\mathbf{k}}^{4} - \mathbf{D}_{\mathbf{k}}^{+} \cdot \mathbf{Q}_{\mathbf{k}}^{a}(\omega) \cdot \mathbf{D}_{\mathbf{k}}^{4} \cdot \mathbf{P}_{\mathbf{k}}^{-}(\omega) \cdot \mathbf{D}_{\mathbf{k}}^{-} \Big) 
\cdot \mathbf{R}_{\mathbf{k}}^{a-}(\omega) \cdot 
\Big( \mathbf{D}_{\mathbf{k}}^{4} - \mathbf{D}_{\mathbf{k}}^{+} \cdot \mathbf{P}_{\mathbf{k}}^{-}(\omega) \cdot \mathbf{D}_{\mathbf{k}}^{4} \cdot \mathbf{Q}_{\mathbf{k}}^{a}(\omega) \cdot \mathbf{D}_{\mathbf{k}}^{-} \Big),
\label{3DGreen33}
\\
\mathbf{G}^{r-1}_{\mathbf{k},44}(\omega)
&= [\mathbf{P}^{+}_{\mathbf{k}}(\omega)]^{-1} - \mathbf{D}_{\mathbf{k}}^{4} \cdot \mathbf{Q}_{\mathbf{k}}^{b-}(\omega) \cdot \mathbf{D}_{\mathbf{k}}^{4} 
\nonumber \\
&- \Big( \mathbf{D}_{\mathbf{k}}^{+} - \mathbf{D}_{\mathbf{k}}^{4} \cdot \mathbf{Q}_{\mathbf{k}}^{b-}(\omega) \cdot \mathbf{D}_{\mathbf{k}}^{+} \cdot \mathbf{P}_{\mathbf{k}}^{+}(\omega) \cdot \mathbf{D}_{\mathbf{k}}^{4} \Big)
\cdot \mathbf{R}_{\mathbf{k}}^{b-}(\omega) \cdot 
\Big( \mathbf{D}_{\mathbf{k}}^{-} - \mathbf{D}_{\mathbf{k}}^{4} \cdot \mathbf{P}_{\mathbf{k}}^{+}(\omega) \cdot \mathbf{D}_{\mathbf{k}}^{-} \cdot \mathbf{Q}_{\mathbf{k}}^{b-}(\omega) \cdot \mathbf{D}_{\mathbf{k}}^{4} \Big),
\label{3DGreen44}
\end{align}
where
\begin{align}
[\mathbf{R}_{\mathbf{k}}^{a\pm}(\omega)]^{-1}
& = [\mathbf{P}^{+}_{\mathbf{k}}(\omega)]^{-1} - \mathbf{D}_{\mathbf{k}}^{\mp} \cdot \mathbf{P}^{-}_{\mathbf{k}}(\omega) \cdot \mathbf{D}_{\mathbf{k}}^{\pm} - \mathbf{D}_{\mathbf{k}}^{\mp} \cdot \mathbf{P}_{\mathbf{k}}^{-}(\omega) \cdot \mathbf{D}_{\mathbf{k}}^{4} \cdot \mathbf{Q}_{\mathbf{k}}^{a}(\omega) \cdot \mathbf{D}_{\mathbf{k}}^{4} \cdot \mathbf{P}_{\mathbf{k}}^{-}(\omega) \cdot \mathbf{D}_{\mathbf{k}}^{\pm},
\label{3DGreenRa}
\\
[\mathbf{R}_{\mathbf{k}}^{b\pm}(\omega)]^{-1}
& = [\mathbf{P}^{-}_{\mathbf{k}}(\omega)]^{-1} - \mathbf{D}_{\mathbf{k}}^{4} \cdot \mathbf{P}^{+}_{\mathbf{k}}(\omega) \cdot \mathbf{D}_{\mathbf{k}}^{4} - \mathbf{D}_{\mathbf{k}}^{4} \cdot \mathbf{P}_{\mathbf{k}}^{+}(\omega) \cdot \mathbf{D}_{\mathbf{k}}^{\pm} \cdot \mathbf{Q}_{\mathbf{k}}^{b\pm}(\omega) \cdot \mathbf{D}_{\mathbf{k}}^{\mp} \cdot \mathbf{P}_{\mathbf{k}}^{+}(\omega) \cdot \mathbf{D}_{\mathbf{k}}^{4},
\label{3DGreenRb}
\end{align}
%and
\begin{align}
[\mathbf{Q}_{\mathbf{k}}^{a}(\omega)]^{-1}
& = [\mathbf{P}^{+}_{\mathbf{k}}(\omega)]^{-1} - \mathbf{D}_{\mathbf{k}}^{4} \cdot \mathbf{P}^{-}_{\mathbf{k}}(\omega) \cdot \mathbf{D}_{\mathbf{k}}^{4},
\label{3DGreenQa}
\\
[\mathbf{Q}_{\mathbf{k}}^{b\pm}(\omega)]^{-1}
& = [\mathbf{P}^{-}_{\mathbf{k}}(\omega)]^{-1} - \mathbf{D}_{\mathbf{k}}^{\mp} \cdot \mathbf{P}^{+}_{\mathbf{k}}(\omega) \cdot \mathbf{D}_{\mathbf{k}}^{\pm}, 
\label{3DGreenQb}
\end{align}
%\begin{align}
%[\mathbb{P}^{r}_{\mathbf{k},\pm}(\omega)]^{-1}
%& = \mathbb{W}_\mathbf{k} \mp \mathbb{D}_{\mathbf{k},3},
%\label{P_pm}
%\end{align}
\end{widetext}
where the Floquet matrices ${\bf P}^{\pm}_{\bf k}(\omega)$ and ${\bf D}^{\pm, 4}_{\bf k}$ are given as
\begin{align}
[{\bf P}^{\pm}_{\mathbf{k}}(\omega)]^{-1, nm} = [\hbar(\omega + n\tilde{\Omega}) &+ i \eta] \delta_{nm} 
- (\epsilon_\mathbf{k}^{nm} \pm d_{\mathbf{k},3}^{nm}),
\label{3DGreenP}
\end{align}
and
\begin{align}
({\bf D}_{\mathbf{k}}^{\pm})^{nm} =& d_{\mathbf{k},\pm}^{nm},
\label{3DGreenD_pm}
\\
({\bf D}_{\mathbf{k}}^{4})^{nm} =& d_{\mathbf{k},4}^{nm} .
\label{3DGreenD_4}
\end{align}
%$(\mathbb{W}_\mathbf{k})_{mn} = [\hbar(\omega + n\Omega) + i \eta] \delta_{mn} - (\epsilon_\mathbf{k})_{mn}$, 
%$(\mathbb{D}_{\mathbf{k},\pm})_{mn} = (d_{\mathbf{k},1}\pm i d_{\mathbf{k},2})_{mn} = (d_{\mathbf{k},\pm})_{mn}$,  
%$(\mathbb{D}_{\mathbf{k},3})_{mn} = (d_{\mathbf{k},3})_{mn}$, and $(\mathbb{D}_{\mathbf{k},4})_{mn} = (d_{\mathbf{k},4})_{mn}$.
with $(\epsilon_{\bf k})_{mn}$ and $({\bf d}_{\bf k})_{mn}$ defined the same as before.

As mentioned in the main text, we are interested in the winding number of the WSL within various 2D subspaces to determine the strong $\mathbb{Z}_2$ invariant.
As an example of such 2D subspaces, let us first consider the 2D subspaces lying parallel to the $k_x\mbox{--}k_y$ plane with the electric field applied along the principal, say, $x$ direction, in which case $\tilde{a}_\parallel=a_\parallel=a_\perp=a$.
Then, the Floquet matrices are given as a function of two conserved momenta $k_\perp (=k_y)$ and $k_z$ as follows:
\begin{widetext}
\begin{align}
\epsilon_\mathbf{k}^{nm}
&= e^{i (n-m) k_\parallel a}
\big\{\big[ C + 2 \tilde{D}_2 (1 - \cos{(k_z a)}) + 2\tilde{D}_1 (2 - \cos{(k_\perp a)}) \big] \delta_{nm} 
- \tilde{D}_1 (\delta_{n,m+1} + \delta_{n,m-1}) \big\},
\\
d_{\mathbf{k},\pm}^{nm} 
&= \mp \frac{1}{2i} e^{i (n-m) k_\parallel a} \tilde{A}_1 \big[ 2 \sin{(k_\perp a)} \delta_{nm} 
\mp \delta_{n,m+1} \pm \delta_{n,m-1} \big],
\\
d_{\mathbf{k},3}^{nm} 
&= e^{i (n-m) k_\parallel a} \big\{ \big[ M + 2 \tilde{B}_2 (1 - \cos{(k_z a)}) + 2\tilde{B}_1 ( 2 - \cos{(k_\perp a)} ) \big] \delta_{nm} 
- \tilde{B}_1 (\delta_{n,m+1} + \delta_{n,m-1}) \big\},
\\
d_{\mathbf{k},4}^{nm} 
&= \tilde{A}_2 \sin{(k_z a)} \delta_{nm},
\end{align}
\end{widetext}
which can be plugged into Eqs.~\eqref{3DGreenP}\mbox{--}\eqref{3DGreenD_4} to compute Eqs.~\eqref{3DGreenRa}\mbox{--}\eqref{3DGreenQb} and subsequently the inverses of the Floquet Green's functions in Eqs.~\eqref{3DGreen11}\mbox{--}\eqref{3DGreen44}, which are then inverted to generate the Floquet Green's functions.

Similarly, we also consider the 2D subspaces lying parallel to the $k_y\mbox{--}k_z$ plane. 
Now, the electric field is applied along the $y$ direction in these 2D subspaces. 
Then, the Floquet matrices are given as a function of two conserved momenta $k_\perp (=k_z)$ and $k_x$ as follows:
\begin{widetext}
\begin{align}
\epsilon_\mathbf{k}^{nm}
& = e^{i (n-m) k_\parallel a} \big\{ \big[ C + 2 \tilde{D}_2 (1 - \cos{(k_\perp a)}) + 2\tilde{D}_1 (2 - \cos{(k_x a)}) \big] \delta_{nm} 
- \tilde{D}_1 (\delta_{n,m+1} + \delta_{n,m-1}) \big\},
\\
d_{\mathbf{k},\pm}^{nm} 
& = \frac{1}{2} e^{i (n-m) k_\parallel a} \tilde{A}_1 \big[ 2 \sin{(k_x a)} \delta_{nm} \pm \delta_{n,m+1} \mp \delta_{n,m-1} \big],
\\
d_{\mathbf{k},3}^{nm} 
& = e^{i (n-m) k_\parallel a} \big\{ \big[ M + 2 \tilde{B}_2 (1 - \cos{(k_\perp a)}) + 2\tilde{B}_1 ( 2 - \cos{(k_x a)} ) \big] \delta_{nm} 
- \tilde{B}_1 (\delta_{n,m+1} + \delta_{n,m-1}) \big\},
\\
d_{\mathbf{k},4}^{nm} 
& = \tilde{A}_2 \sin{(k_\perp a)} \delta_{nm}, 
\end{align}
\end{widetext}
which can be used similarly to generate the Floquet Green's functions.

\end{document}